\documentstyle[12pt]{article}

\begin{document}
\title{Physics at LHC}
\author{N.V.Krasnikov and V.A.Matveev \\
INR RAS, Moscow 117312}

\date{February, 1997}
\maketitle
\begin{abstract}
We review the physics to be investigated at LHC. We also describe the 
main parameters of CMS and ATLAS detectors.

\end{abstract}

\newpage

{\bf Content}

1. {\bf Introduction}

2. {\bf Detectors}

2.1 CMS detector

2.2 ATLAS detector

3. {\bf Physics at LHC}

3.1 Parton model

3.2 Physics within SM

3.2.1 Top-quark physics

3.2.2 Higgs boson in Weinberg-Salam model

3.2.3 Search for standard Higgs boson at CMS

3.2.4 B-physics in SM

3.2.5 B-physics at CMS

3.2.6 Heavy ion physics

3.3 Supersymmetry search within MSSM

3.3.1 MSSM model

3.3.2 The search for SUSY Higgs bosons

3.3.3 Squark and gluino search

3.3.4 Neutralino and chargino search

3.3.5 Sleptons search

3.4 The search for physics beyond SM and MSSM

4. {\bf Conclusion}

\newpage

\newpage
\section{Introduction}

The scientific program at LHC(Large Hadron Collider) \cite{1,2} which
will be the biggest particle accelerator complex ever built in the
World consists in many goals. Among them there are two supergoals:

a. Higgs boson discovery in standard electroweak Weinberg-Salam model.

b. Supersymmetry discovery.

LHC will accelerate two proton beams with the total energy $\sqrt{s} =
14$Tev. At low luminosity stage (first two-three years of operation)
the luminosity is planned to be $L_{low} = 10^{33}cm^{-2}s^{-1}$
with total luminosity $L_{tot} = 10^{4}pb^{-1}$ per year.
At high luminosity stage the luminosity is planned to be 
$L_{high} = 10^{34}cm^{-2}s^{-1}$ with total luminosity 
$L_{tot} = 10^{5}pb^{-1}$ per year. The LHC will start to
work in 2005 year. There are a lot of lines for research for LHC:

a.  Higgs boson discovery in standard electroweak Weinberg-Salam model.

b.  Supersymmetry discovery.

c. B-physics.

d. Heavy ion physics.

e. Top quark physics.

f. Standard physics (QCD, electroweak interactions).

g. The search for new physics beyond minimal supersymmetric model and
   Weinberg-Salam model.

There are planned to be two big detectors at LHC - CMS(Compact Muon Solenoid)
and ATLAS.

In this paper we describe the main parameters of CMS and ATLAS detectors
and review the main physics to be investigated at LHC.\footnote{ To be precise,
we review the main physics to be investigated at CMS}

\section{Detectors}

\subsection{CMS detector}

\paragraph{Subdetectors.}

CMS detector consists of the following subdetectors:

1.  Tracker barrel detector(TB).

2.  Tracker forward detector(TF).

3.  ECAL(electromagnetic calorimeter) barrel detector(EB).

4.  ECAL forward(endcap) detector(EF).

5.  Preshower in front of barrel ECAL(SB).

6.  Preshower in front of forward ECAL(SF).

7.  HCAL(hadronic calorimeter) barrel detector(HB).

8.  HCAL forward(endcap) detector(HF).

9.  Very forward calorimeter(VF).

10.  Barrel muon station(MS).

11.  Forward muon station(MF).

\paragraph{Basic goals and design considerations.}

One of the most important tasks for LHC is the quest for
the origin of the spontaneous symmetry-breaking mechanism
in the electroweak sector of the standard model(SM).
Namely, all the renormalizable models of electroweak
interactions are based on the use of the gauge symmetry
breaking. As a consequence of the electroweak symmetry breaking and
the renormalizability of the theory there must be neutral
scalar particle(Higgs boson) in the
spectrum. The existing LEP1 bound on the Higgs boson mass in SM is
$m_{h} \geq 64$ Gev. LEP2 with the full energy
$\sqrt{s} = (190 -195)$ Gev will be able to discover the Higgs
boson in SM with a mass up to $m_{h} \leq (90 - 95)$ Gev.
Theoretical bound based on tree
level unitarity gives upper bound $m_{h} \le 1$Tev for the Higgs
boson mass. Similar bound $m_{h} \le (700 - 800)$ Gev  give lattice based
estimates.  In minimal supersymmetric extension of SM(MSSM) the
lightest Higgs boson has to be relatively light $m_{h} \le 120$ Gev.
Other nonminimal supersymmetric models of electroweak interactions
typically predict relatively light lightest Higgs boson $m_{h} \le
(160 - 180)$ Gev. So the discovery of the Higgs boson will be the
check of the spontaneous symmetry breaking and the renormalizability
of the theory and there are no doubts that it is the supergoal number
1 for LHC. The Higgs search is therefore used as a first benchmark
for the detector optimisation for both CMS and ATLAS. For the SM
Higgs boson, the detector has to be sensitive to the
following processes in order to cover the full mass range
above the expected LEP2 discovery limit of $(90 - 95)$ Gev:

A. $h \rightarrow \gamma \gamma$ mass range  90 Gev$ \le m_{h} \le $
150 Gev.

B. $h \rightarrow b\bar{b}$ from $Wh, Zh, t\bar{t}h$ using
$l^{\pm}(l^{\pm} = e^{\pm}$ or $\mu^{\pm})$- tag and
b-tagging in the mass range 80 Gev$ \le m_{h} \le $100 Gev.

C. $h \rightarrow ZZ^{*} \rightarrow 4l^{\pm}$ for mass range
$130 Gev \le m_{h} \le 2m_{Z}$.

D. $h \rightarrow ZZ \rightarrow 4l^{\pm}, 2l^{\pm}2\nu$ for the
mass range $m_{h} \ge 2m_{Z}$.

E. $h \rightarrow WW, ZZ \rightarrow l^{\pm}\nu$ 2 jets, $2l^{\pm}$
2 jets, using tagging of forward jets for $m_{h}$ up to $\; 1$ Tev.

In minimal supersymmetric extension of the standard model(MSSM)
there is a family of Higgs particles $(H^{\pm}, h, H$ and $A$).
So in addition to  the standard Higgs boson signatures the
MSSM Higgs searches are based on the following processes:

F. $ A \rightarrow \tau^{+}\tau^{-} \rightarrow e \mu$ plus $\nu ' s$,
or $A \rightarrow \tau^{+} \tau^{-} \rightarrow l^{\pm}$
plus hadrons plus $\nu 's$.

G. $H^{\pm} \rightarrow \tau^{\pm} \nu$ from $t\bar{t} \rightarrow
H^{\pm}W^{\mp}b\bar{b}$ and $H^{\pm} \rightarrow 2$ jets, using a
$l^{\pm}$- tag and $b$-tagging.

The observable cross sections for most of those processes are small
$(1 - 100) pb$ over a large part of the mass range. So it is
necessary to work at high luminosity and to maximize
the detectable rates above backgrounds by high-resolution
measurements of electrons, muons and photons.

For the $H^{\pm}$ and $A$ signatures in the case of the MSSM,
high  performance detector capabilities are required in addition for
the measurements which are expected to be best achieved at initial
luminosities with a low level of overlapping events, namely secondary
vertex detection for $\tau$-leptons  and b-quarks, and high resolution
calorimetry for jets and missing transverse energy $E^{miss}_{T}$.

The second supergoal of the LHC project is the supersymmetry discovery,
i.e. the detection of superparticles. Here the main
signature are the missing transverse energy events which are
the consequence of undetected lightest stable supersymmetric
particles LSP predicted in supersymmetric models with R-parity
conservation. Therefore it is necessary to set stringent requirements
for the hermeticity and $E^{miss}_{T}$ capability of the detector.
Also the search for new physics different from supersymmetry
(new gauge bosons $W^{'}$ and $Z^{'}$, new Higgs bosons with big
Yukawa couplings etc.) at LHC requires high resolution lepton
measurements and charge identification  even in
the $p_{T}$ range of a few Tev. Other possible signature of new
physics(compositeness) can be provided by very high $p_{T}$ jet
measurements. An important task of LHC is the study of b- and
t-physics. Even at low luminosities the LHC will be a high rate
beauty- and top-quark factory. The main emphasis in B-physics
is the precise measurement of CP-violation in the $B^{0}_{d}$
system and the determination of the Kobayashi-Maskawa angles. Besides,
investigations of $B\bar{B}$ mixing in the $B^{0}_{S}$ system, rare B
decays are also very important. Precise secondary vertex determination,
full reconstruction of final states with relatively low-$p_{T}$
particles, an example being $B^{0}_{d} \rightarrow J/\Psi K^{0}_{S}$
followed by $J/\Psi \rightarrow l^{+}l^{-}$ and $K^{0}_{S}
\rightarrow \pi^{+}\pi^{-}$, and low-$p_{T}$ lepton first-level
triggering capability are all necessary. In addition to running
as a proton-proton collider, LHC will be used to collide heavy
ions at a centre of mass energy 5.5 Tev per nucleon pair.
The formation of quark-gluon plasma in the heavy ion colisions
is predicted to be signalled by a strong suppression of
$\Upsilon^{'}$ and $\Upsilon^{''}$ production relative to
$\Upsilon$ production when compared with pp collisions.
The CMS and ATLAS detectors will be
used to detect low momentum muons produced in heavy ion collisions
and reconstruct $\Upsilon$, $\Upsilon^{'}$ and $\Upsilon^{''}$ meson
production. Therefore the basic design considerations for both ATLAS
and CMS are the following:

1. very good electromagnetic calorimetry for electron and photon
identification and measurements,

2. good hermetic jet and missing $E_{T}$-calorimetry,

3. efficient tracking at high luminosity for lepton momentum
measurements, for b-quark tagging, and for enhanced electron and photon
identification, as well as tau and heavy-flavour vertexing and
reconstruction capability of some B decay final states at lower
luminosity,

4. stand-alone, precision, muon-momentum measurement up to highest
luminosity, and very low-$p_{T}$ trigger capability at lower luminosity,

5. large acceptance in $\eta$ coverage.

\paragraph{Brief description of CMS subdetectors}

\subparagraph{Tracker.}

The design goal of the cental tracking system is to reconstruct
isolated high $p_{T}$ tracks with an efficiency better than $95$
percent, and high $p_{T}$ tracks within jets with an efficiency of
better than 90 percent over the rapidity $|\eta| \le 2.6$. The
momentum resolution required for isolated charged leptons in the
central rapidity region is $\frac{\delta p_{T}}{p_{T}} =  0.1p_{T}$
$(p_{T}$ in Tev). This will allow the measurement of the lepton
charge up to $p_{T} = 2$ Tev. It is also very important for tracking
system to perform efficient b- and $\tau$-tagging.  The tracker
system consists of silicon pixels, silicon and gas microstrip
detectors(MSGS) which provide precision momentum measurements and
ensure efficient pattern of recognition even at the highest
luminosity. A silicon pixel detectors consist of two barrel layers
and three endcap layers and it is placed close to the beam pipe with
the tasks of:

a. assisting in pattern recognition by providing two or three
true space points per track over the full rapidity range in the
main tracker,

b. improving the impact parameter resolution for b-tagging,

c. allowing 3-dimensional vertex reconstruction by providing a much
improved Z-resolution in the barrel part.

The silicon microstrip detector is required to have a powerful vertex finding
capability in the transverse plane over a large momentum range for b-tagging
and heavy quark physics and must be able to distinguish different interaction
vertices at high luminosity. The CMS silicon microstrip detector is
subdivided into barrel and forward parts, meeting at $|\eta| = 1.8 (\eta
\equiv -\ln(\tan(\frac{\theta}{2}))$, provided at least 3 measuring
points on each track for $|\eta| \le 2.6$. The microstrip gas
chambers provide a minimum of 7 hits for high $p_{T}$ tracks. The
track finding efficiency in the tracker is 98 percent for $p_{T} \ge
5$ Gev. The charged particle momentum resolution depends on the
$\eta$ and $p_{T}$ of charged particle and for $p_{T} = 100$ Gev and
$|\eta| \le 1.75$ it is around 2 percent. Impact parameter resolution
also depends on $p_{T}$ and $\eta$ and for 10 Gev$ \le p_{T} \le 100$
 Gev and $|\eta| \le 1.3$ in transverse plane it is around 100
$\mu m$. The b-tagging efficiency from $\bar{t}t$ decays is supposed to be
better than 30 percent. A significant impact parameter can be used to
tag $\tau$-leptons. It could be useful in searches such as SUSY Higgs
boson decays $A,H,h \rightarrow
\tau\tau \rightarrow e + \mu + X$(or $l +
hadrons)$.  These leptons(hadrons) originate from secondary $(\tau)$
vertices while in the backgrounds from $\bar{t}t \rightarrow Wb +
W\bar{b} \rightarrow e + \mu + X$ and
$WW \rightarrow e + \mu + X$ they originate
from the primary vertex. It is possible to have the
efficiency for the signal $\approx 50$ percent while for the
background channels it is $\approx 3$ percent.

\subparagraph{ECAL.}

The barrel part of the electromagnetic calorimeter covers the rapidity
intervals $|\eta| \le 1.56$. The endcaps cover the intervals
$1.65 \le |\eta| \le 2.61$. The gaps between the barrel and the
endcaps are used to route the services of the tracker and preshower
detectors.  The barrel granularity is 432 fold in $\phi$ and
$108 \times 2$-fold in $\eta$.  A very good intrinsic energy
resolution given by
\begin{equation}
\frac{\sigma}{E} =
\frac{0.02}{\sqrt{E}} \oplus 0.005 \oplus \frac{0.2}{E}
\end{equation}
is assumed to be for electrons and photons with a $PbWO_{4}$ crystal
ECAL. The physics process that imposes the strictest performance
requirements on the electromagnetic calorimeter is the intermediate
mass Higgs decaying into two photons. The main goal here is to obtain
very good di-photon mass resolution. The mass resolution has terms
that depend on the resolution in energy $(E_{1}, E_{2})$ and the two
photon angular separation $(\theta)$ and it is given by
\begin{equation}
\frac{\sigma_{M}}{M} = \frac{1}{2}[\frac{\sigma_{E_{1}}}{E_1} \oplus
\frac{\sigma_{E_2}}{E_2} \oplus
\frac{\sigma_{\theta}}{(\tan(\frac{\theta}{2})}] \,,
\end{equation}
where $\oplus$ denotes a quadratic sum, $E$ is in Gev
and $\theta$ is in radians. For the Higgs two-photon decay at LHC the angular
term in the mass resolution can become important, so it is necessary to
measure the direction of the photons using the information from the
calorimeter alone. In the barrel region $|\eta| \le 1.56$ angular
resolution is supposed to be $\sigma_{\theta} \le \frac{50
mrad}{\sqrt{E}}$.  Estimates give the following di-photon mass
resolution for $h \rightarrow \gamma \gamma$ channel $(m_{h} = 100$
 Gev):

$\delta m_{\gamma\gamma} = 475$ Mev (Low luminosity
$L = 10^{33}cm^{-2}s^{-1}$),

$\delta m_{\gamma\gamma} = 775$ Mev
(High luminosity $L =
10^{34}cm^{-2}s^{-1}$).

\subparagraph{HCAL.}

The hadron calorimeter surrounds the electromagnetic calorimeter and acts in
conjuction with it to measure the energies and directions of particle jets,
and to provide hermetic coverage for measurement the transverse energy. The
pseudorapidity range $( |\eta| \le 3)$ is covered by the barrel and
endcap hadron calorimeters which sit inside the $4T$ field of CMS
solenoid. In the central region around $\eta = 0$ a hadron shower
'tail catcher' is installed outside the solenoid coil to ensure
adequate sampling depth. The active elements of the barrel and endcap
hadron calorimeter consist of plastic scintillator tiles with
wave length-shifting fibre readout. The pseudorapidity range $( 3.0
\le \eta \le 5.0)$ is covered by a separate very forward
calorimeter. The hadron calorimeter must have good hermeticity, good
transverse granularity, moderate energy resolution and sufficient
depth for hadron shower containment. The physics programme requires
good hadron resolution and segmentation to detect narrow states
decaying into pairs of jets. The di-jet mass resolution includes
contributions from physics effects such as fragmentation as well as
detector effects such as angular and energy resolution. The energy
resolution is assumed to be:
\begin{equation}
\frac{\Delta E}{E} = \frac{0.6}{\sqrt{E}} \oplus 0.03
\end{equation}
for $|\eta| \le 1.5$ and segmentation $\Delta \eta \times \Delta \Phi
= 0.1 \times 0.1$.

The di-jet mass resolution is approximately the following:

1. (10 -15) percent for 50 Gev$ \le p_{T} \le 60$ Gev and $m_{ij} =m_Z$.

2. (5 - 10) percent for 500 Gev$ \le p_{T} \le $600 Gev and $m_{ij} =m_Z$.

The expected energy resolution for jets in the very forward 
calorimeter is parametrized by:
\begin{equation}
\frac{\sigma_{E_{jet}}}{E_{jet}} = \frac{1.28 \pm 0.1}{\sqrt{E_{jet}}}
\oplus (0.02 \pm 0.01)\,.
\end{equation}
The expected missing transverse energy resolution in the CMS detector with
very forward $2.5 \le \eta \le 4.7$ coverage is
\begin{equation}
\frac{\sigma_t}{\sum E_t} = \frac{0.55}{\sqrt{\sum E_t}} \,,
\end{equation}
($E_t$ in Gev). In the absence of the very forward calorimeter, the missing
transverse energy resolution would be nearly three times worse.

\subparagraph{Muon system.}

At the LHC the effective detection of muons from Higgs bosons, W, Z and
$t\bar{t}$ decays requires coverage over a large rapidity interval. Muons
from pp collisions are expected to provide clean signatures for a wide
range of new physics processes. Many of these processes are expected
to be rare and will require the highest luminosity. The goal of the muon
detector is to identify these muons and to provide a precision measurement
of their momenta from a few Gev to a few Tev. The barrel detector covers
the region $|\eta| \le 1.3$. The endcap detector covers the region
$1.3 \le |\eta| \le 2.4$. The muon detector should filfil three
basic tasks: muon identification, trigger and momentum measurement.
The muon detector is placed behind ECAL and the coil. It consists of
four muon stations interleaved with the iron return yoke plates. The
magnetic flux in the iron provides the  possibility of an independent
momentum measurement.  The barrel muon detector is based on a system
of 240 chambers of drift tubes arranged in four concentric stations.
In the endcap regions, the muon detector comprises four muon
stations. The muon detector has the following functionality and
performance:

1. Geometric coverage: pseudorapidity coverage up to $|\eta| =2.4$ with
the minimum possible acceptance loses due to gaps and dead areas.

2. Transverse momentum resolution for the muon detector alone for
$0 \le |\eta| \le 2$ : $\frac{\Delta p_{T}}{p_{T}} = 0.06 - 0.1$ for
$p_{T} = 10$ Gev, $0.07 - 0.2$ for $p_{T} = 100$ Gev and $0.15 -
0.35$ for $p_{T} = 1$ Tev.

3. Transverse momentum resolution after matching with central detector
for $0 \le |\eta| \le 2$ : $\frac{\Delta p_{T}}{p_{T}} = 0.005 -
0.01$ for $p_{T} = 10$ Gev, $0.015 - 0.05$ for $p_{T} = 100$ Gev
and $0.05 - 0.2$ for $p_{T} = 1$ Tev.

4. Charge assignment: correct at 99 percent confidence level up to
  $p_{T} = 7$ Tev for the full $\eta$ coverage.

5. Muon trigger: precise muon chambers and fast dedicated detectors provide
 a trigger with $p_{T}$ thresholds from a few Gev up to 100 Gev.

A schematic view of the CMS detector is shown in Fig.1.1, Fig.1.2 and
Fig.1.3 .

\subsection{ATLAS detector}

The design of the ATLAS detector is similar to CMS detector. It also consists
of inner detector(tracker), electromagnetic calorimeter, hadron calorimeter
and muon spectrometer. Here we briefly describe the main parameters of
the ATLAS subdetectors.

\paragraph{Inner detector.}

The main parameters of the ATLAS inner detector are:

1. Tracking coverage over the pseudorapidity range $|\eta| \le 2.5$.

2. Momentum resolution of $\frac{\Delta p_{T}}{p_T} \le 0.3$ at
$p_T = 500$ Gev for $|\eta| \le 2$ and no worse than 50 percent for
$|\eta| = 2.5$.

3. Polar-angle resolution  of $\le 2$ mrad.

4. Tracking efficiency of $\geq 95$ percent over the full coverage for
isolated tracks with $p_{T} \geq 5$ Gev, with fake-track rates less
than 1 percent of signal rates.

5. Tagging of b jets with an efficiency $\ge 30$ percent at the
highest luminosity, with a rejection $\ge 10$ against non b- hadronic
jets.

6. For initial lower-luminosity running the ability to reconstruct secondary
vertices from b and $\tau$ decays and charged tracks from primary
vertices and from secondary decay vertices of short-lived particles
with $\ge 95$ percent efficiency for $p_T \ge 0.5$ Gev over the full
coverage.

\paragraph{ECAL.}

The energy resolution is of $ \frac{\Delta E}{E} =
\frac{0.1}{\sqrt{E}} \oplus 0.007$ for $|\eta| \le 2.5$. Diphoton
mass resolution is estimated to be 1.4 Gev for Higgs boson mass $m_h
= 100$ Gev for $L = 10^{34}cm^{-2}s^{-1}$ (for CMS the diphoton mass
resolution is 775 Mev).

\paragraph{HCAL.}

Jet energy resolution is of $\frac{\Delta E}{E} = \frac{0.5}{\sqrt{E}}
\oplus 0.03$ for jets and a segmentation of $\Delta \eta \times
\Delta \Phi = 0.1 \times 0.1$ for $|\eta| \le 3$ and
$\frac{\Delta E}{E} = \frac{1}{\sqrt{E}} \oplus 0.1$ and a segmentation
of $\Delta \eta \times \Delta \Phi = 0.1 \times 0.1$ for very forward
calorimeter $3 \le |\eta| \le 5$.

\paragraph{Muon spectrometer.}

The muon momentum resolution is of $\frac{\Delta p_{T}}{p_T} = 0.02
(p_T = 20$ Gev), $\frac{\Delta p_T}{p_T} = 0.02(p_T = 100$ Gev),
$\frac{\Delta p_{T}}{p_T} = 0.08(p_T = 1$ Tev) for $|\eta| \le 3$.

\section{Physics at LHC}

\subsection{Parton model}

A high-energy proton beam may  be regarded as an unseparated
beam of quarks, antiquarks and gluons. For the hard-scattering
phenomena that are the principal interest of LHC the cross
section for the hadronic reaction
\begin{equation}
a + b \rightarrow c + anything
\end{equation}
is given in parton model by the formula \cite{3,4}
\begin{equation}
d\sigma(a + b \rightarrow c + X) = \sum_{partons:
   i,j}f_i^{(a)}f_j^{(b)}d\hat{\sigma}(i + j \rightarrow c + X^{'})\,,
\end{equation}
where $f^{(a)}_{i}$ is the probability of finding constituent $i$ in
hadron $a$ and $\hat{\sigma}(i + j  \rightarrow c + X^{'})$ is the
cross section for the elementary process leading to the desired
final state. This picture of hadron collisions in many cases
provides a reliable estimate of cross sections. Two ingredients
are required to compute cross sections: the elementary cross
sections and the parton distributions. It is straightforward to
calculate the elementary cross sections, at least at low orders in
perturbation theory.  At given scale, the parton distributions can
be measured in deep inelastic lepton-hadron scattering. The
evolution of these distributions to larger momentum scales is
described by perturbative QCD.

As it has been mentioned before
the main idea of parton model is to regard a high energy proton
   as a collection of quasifree partons which share its momentum.
   Thus we consider a proton with momentum $P$ as being made of
   partons carrying longitudinal momenta $x_iP$, where the momentum
   fractions $x_i$ satisfy
   \begin{equation}
   0 \leq x_i \leq 1
   \end{equation}
   and
   \begin{equation}
   \sum_{partons \, i} x_i = 1\,.
   \end{equation}
   The cross section for the reaction (6) is given by
   \begin{equation}
   d\sigma(a + b \rightarrow c + X) = \sum_{ij} f_i^{(a)}(x_a)
   f_j^{(b)}(x_b)d\hat{\sigma}(i + j \rightarrow c + X^{'})\,,
   \end{equation}
   where $f^{(a)}_{i}(x)$ is the number distribution of partons of
   species $i$. The summation runs over all contributing parton
   configurations. Denote the invariant mass of the $i-j$ system as
   \begin{equation}
   \sqrt{\hat{s}} = \sqrt{s\tau}
   \end{equation}
   and its longitudinal momentum in the hadron-hadron c.m. by
   \begin{equation}
   p =x\sqrt{s}/2 \,.
   \end{equation}
   The kinematic variables $x_{a,b}$ of the elementary process are
   related to those of the hadronic process by
   \begin{equation}
   x_{a,b} = \frac{1}{2}[(x^2 + 4\tau)^{1/2} \pm x] \,.
   \end{equation}
   The parton momentum fractions satisfy the relations
   \begin{equation}
   x_ax_b = \tau \,,
   \end{equation}
   \begin{equation}
   x_a - x_b = x \,.
   \end{equation}

The elementary parton model as described here does not take into
account QCD strong interaction effects. The most important modification
of naive parton model is due to QCD corrections to the parton
distributions. In leading logarithmic approximation \cite{5} these
corrections are process independent and can be incorporated by the
replacement
\begin{equation}
f_{i}^{(a)}(x_a) \rightarrow f_{i}^{(a)}(x_a,Q^2)\,.
\end{equation}
There is some ambiguity in the choice of scale $Q^2$ for a particular
process. It should be of the order of the subenergy $Q^2 \approx
\hat{s}$. The knowledge of the next to leading corrections allows to
fix the scale $Q^2$. Usually we shall neglect higher-order
QCD corrections. Experience shows that in many cases an account
of higher order corrections increases the value of cross sections by
factor $1.5 - 2$ \cite{5}.

To calculate production cross sections for a hadron collider we have
to know parton distributions as functions of the scaling variable
x and $Q^2$. For the study of a process with characteristic mass M,
the parton distributions must be known for $Q^2 \approx M^2$ and
$x \geq M^2/s $. The typical momentum fraction contributing
to such a process is $x \approx M/\sqrt{s}$. We shall be interested
in masses $M \geq O(100)$ Gev, so for LHC $x \geq O(10^{-2})$ and
$Q^2 \geq O(10^4)$ $Gev^2$. Although the distributions have not been
measured at such values of $Q^2$ it is possible to obtain
them using the Altarelli-Parisi equation and the parton distributions
at some scale $Q^2_0$. It is convenient to parametrize the
distributions in a valence plus sea plus gluon form. The proton
contains:

up quarks: $u_v(x,Q^2) + u_s(x,Q^2)$,

down quarks: $d_v(x,Q^2) + d_s(x,Q^2)$,

up antiquarks $u_s(x,Q^2)$,

down antiquarks: $d_s(x,Q^2)$,

strange, charm, bottom and top quarks and antiquarks: $q_s(x,Q^2)$,

gluons: $G(x,Q^2)$.

The flavour quantum numbers of the proton are carried by the valence
quarks. Those distributions must therefore satisfy the number sum
rules
\begin{equation}
\int_{0}^{1} dx\,u_v(x,Q^2) = 2\,,
\end{equation}
\begin{equation}
\int_{0}^{1}dx\,d_v(x,Q^2) = 1 \,.
\end{equation}
The parton distributions are also constrained by the momentum sum
rule
\begin{equation}
\int_{0}^{1} dx\,x[u_v + d_v + G +2(u_s + d_s + s_s + c_s + b_s
	+t_s)] = 1 \,.
\end{equation}
For protons, many sets of parton distributions exist on the market.
These are obtained by fits to experimental data, constrained so that
the $Q^2$ dependence is in accordance with the standard QCD evolution
equations. At present the most popular set of parton distributions is
CTEQ2L \cite{6}. Also available in PYTHIA program \cite{7} are
EHLQ, DO, CTEQ2M, CTEQ2MS, CTEQ2MF, CTEQ2ML and CTEQ2D sets
of parton distributions.
The Altarelli-Parisi equation for nonsinglet  distributions reads
\begin{eqnarray}
&&\frac{dp(x,Q^2)}{d\ln{Q^2}} = \frac{2\alpha_s(Q^2)}{3\pi}\int_{x}^{1}
dz[\frac{(1+z^2)p(y,Q^2)- 2p(x,Q^2)}{1-z}] \\ \nonumber
&&+ \frac{\alpha_s(Q^2)}{\pi}[1 + \frac{4\ln{(1-x)}}{3}]p(x,Q^2)\,,
\end{eqnarray}
where
\begin{equation}
p(x,Q^2) = xu_v(x,Q^2)\;or\;xd_v(x,Q^2)
\end{equation}
and $y = x/z$. The evolution of the gluon momentum distribution
\begin{equation}
g(x,Q^2) = xG(x,Q^2)
\end{equation}
is given by
\begin{eqnarray}
&&\frac{dg(x,Q^2)}{d\ln{Q^2}} = \frac{\alpha_{s}(Q^2)}{\pi}
\int_{x}^{1}dz[ \frac{3[zg(y,Q^2)-g(x,Q^2)]}{1-z} + \\ \nonumber
&&\frac{3(1-z)(1+z^2)g(y,Q^2)}{z} +
\frac{2}{3}\frac{1+(1-z)^2}{z} \\ \nonumber
&&\sum_{flavors: q}y[q_v(y,Q^2) + 2q_s(y,Q^2)] +
\frac{\alpha_s(Q^2)}{\pi}[\frac{11}{4} \\ \nonumber
&&- \frac{N_f}{6} +
3\ln{(1-x)}]g(x,Q^2)\,,
\end{eqnarray}
where $N_f$ is the number of flavours participating in the evolution
at $Q^2$. The evolution of the momentum distributions of the light
sea quarks
\begin{equation}
l(x,Q^2) = xu_s(x,Q^2)\; or\;xd_s(x,Q^2)\;or\;xs_s(x,Q^2)
\end{equation}
is described by
\begin{eqnarray}
&&\frac{dl(x,Q^2)}{d\ln{Q^2}} = \frac{2\alpha_s(Q^2)}{3\pi}\int^1_x
dz[\frac{(1 + z^2)l(y,Q^2) -2l(x,Q^2)}{1-z}  + \\ \nonumber
&&\frac{3}{8}[z^2 +
(1 - z^2)]g(y,Q^2)] + \frac{\alpha_s(Q^2)}{\pi}[1 + \frac{4}{3}
\ln{(1-x)}]l(x,Q^2) \,.
\end{eqnarray}
For the evolution of the momentum distributions of heavy sea quarks
\begin{equation}
h(x,Q^2) =xc_s(x,Q^2)\;or \; xb_s(x,Q^2)\; or \; xt_s(x,Q^2)
\end{equation}
the evolution equation reads \cite{8}
\begin{eqnarray}
&&\frac{dh(x,Q^2)}{d\ln{Q^2}} = \frac{2\alpha_s(Q^2)}{3\pi} \int_x^1
dz[\frac{(1 + z^2)h(y,Q^2) -2h(x,Q^2)}{1-z} + \\ \nonumber
&&\frac{3}{4\beta}
[\frac{1}{2} - z(1 - z) +
\frac{M_q^2}{Q^2}\frac{(3-4z)z}{1-z} -
	 \frac{16M_q^4z^2}{Q^4}]g(y,Q^2)
- \frac{3M_q^2}{2Q^2}[z(1 - 3z) + \frac{4M_q^2z^2}{Q^2}] \\ \nonumber
&&\ln(\frac{1 +\beta}{1-\beta})g(y,Q^2)]\theta(\beta^{2})
+ \frac{\alpha_s(Q^2)}{\pi}[1 + \ln(1 - x)]h(x,Q^2)\,,
\end{eqnarray}
where $M_q$ is the heavy quark mass and
\begin{equation}
\beta = [1 -4M^2_q/Q^2(1 - z)]^{1/2}\,.
\end{equation}
The formula for the running effective strong coupling constant for
massless quarks in the leading log approximation has the form
\begin{equation}
\alpha_s(Q^2) = \frac{12\pi}{(33 - 12N_f)\ln(Q^2/\Lambda^{2})} \,.
\end{equation}
We shall use the formula (29) for $N_f = 4$. For the case of massive
quarks the generalization of the formula (29) takes the form
\begin{equation}
1/\alpha_s(Q^2) = \frac{25}{12\pi}\ln{(Q^2/\Lambda^2)} -
\frac{1}{6\pi}\sum_{i =b,t,...}\theta(Q^2 -
16M^2_i)\ln{(Q^2/16M^2_i)}\,.
\end{equation}

The parton distributions are necessary to calculate differential
and total cross sections. It appears that the cross sections depend
on some particular combination of structure functions, parton -
parton luminosities. Namely, the differential cross section for the
reaction
\begin{equation}
a + b \rightarrow \alpha + anything
\end{equation}
is given by
\begin{equation}
\frac{d\sigma}{d\tau}(a + b \rightarrow \alpha + X) =
\sum_{ij}\frac{dL_{ij}}{d\tau}\hat{\sigma}(ij \rightarrow \alpha) \,,
\end{equation}
where $\hat{\sigma}(ij \rightarrow \alpha)$ is the cross section
for the corresponding elementary process, $\tau = \hat{s}/s$ and
\begin{equation}
\frac{dL_{ij}}{d\tau} = \frac{1}{1 + \delta_{ij}}\int^1_{\tau}dx
[f^{(a)}_i(x)f^{(b)}_j(\tau/x) + f^{(a)}_j(x)f^{(b)}_i(\tau/x)]/x \,.
\end{equation}
Here $f^{(a)}_i(x)$ is the number distribution of partons of
species i carrying momentum fraction $x$ of hadron a. The
hard-scattering processes that determine the study for
interesting physics at LHC have a common asymptotic form prescribed
by dimensional analysis
\begin{equation}
\sigma(\hat{s}) = c/\hat{s} \,.
\end{equation}
For a strong-interaction process, such as jet pair production, $c$ is
typically of order $(\alpha_s/\pi)^2$. For a typical electroweak
process such as lepton pair production, $c$ is approximately
$(\alpha/\pi)^2$. Resonance production cross sections are
proportional to $\tau$. Therefore the quantity $\frac{\tau}{\hat{s}}
\frac{dL}{d\tau}$, which has dimension of the cross section,
provides a useful measure of the reach of  hadron collider at
given energy and hadron-hadron luminosity.

\subsection{Physics within SM}

\paragraph{3.2.1 Top-quark physics}

\subparagraph{ }

Even at low initial luminosities of $10^{32}cm^{-2}s^{-1}$ approximately
6000 $t\bar{t}$ pairs would be produced per day at $m_{t} = 170$ Gev,
yielding about 100 reconstructed $t\bar{t} \rightarrow (l\nu b)(jjb)$
decays per day and about 10 clean isolated $e\mu$ pairs per day. The
$t \rightarrow jjb$ decays provide an abundant event sample, which
allows direct reconstruction of $m_t$, through the invariant mass of
3-jet system. The conclusion is that the ultimate accuracy of $\pm 2$ Gev
may be achieved for $m_t = 170$ Gev \cite{2}.

Multilepton events from top-quark decays can also be used to extract an
accurate measurement of $m_{t}$. The most promising method is to use
2 leptons from the same top-quark decay. For an integrated luminosity
of $10^{4}pb^{-1}$, the expected statistical uncertainty on the measurement
of $m_t$, using this method, is approximately $\pm 0.5$ Gev for
$m_t = 170$ Gev. The overall systematic uncertainty due to fragmentation
effects will very likely to be less than $\pm 2$ Gev. The conclusion
is that the accuracy of better than $\pm 2$ Gev may be achieved in this
channel.

\paragraph{3.2.2 Higgs boson in Weinberg-Salam model}

\subparagraph{  }

The standard Weinberg-Salam model is the renormalizable model of strong
and electroweak interactions. It has the gauge group
$SU_c(3) \otimes SU(2)_L \otimes U(1)$ and the minimal Higgs
structure consisting of one complex doublet of scalar particles. The
spontaneous electroweak symmetry breaking $SU_c(3) \otimes SU_L(2)
\otimes U(1) \rightarrow SU_c(3) \otimes U(1)$ due to nonzero vacuum
expectation value of the Higgs doublet provides the simplest
realization of the Higgs mechanism \cite{9} which generates masses for
$W^{\pm}$, $Z$ gauge bosons and masses to quarks and leptons.  In
this approach, the Goldstone bosons are generated by dynamics of
elementary scalar fields and precisely one neutral Higgs scalar (the
Higgs boson) remains in the physical spectrum. The lagrangian of
Weinberg-Salam model consists of several pieces \cite{10}:
\begin{equation}
L_{WS} = L_{YM} + L_{HYM} + L_{SH} + L_{f} + L_{Yuk}\,.
\end{equation}
Here $L_{YM}$ is the Yang-Mills lagrangian without matter fields
\begin{equation}
L_{YM} =
-\frac{1}{4}F^i_{\mu\nu}(W)F^{\mu\nu}_i(W) - \frac{1}{4}
F^{\mu\nu}(W^0)F_{\mu\nu}(W^0) -
\frac{1}{4}F^a_{\mu\nu}(G)F_a^{\mu\nu}(G)\,,
\end{equation}
where $F^i_{\mu\nu}(W)$, $F^a_{\mu\nu}(G)$, $F_{\mu\nu}(W^0)$ are
given by
\begin{equation}
F^i_{\mu\nu}(W) = \partial_{\mu} W^i_{\nu} - \partial_{\nu}W^i_{\mu}
+g_2\epsilon^{ijk}W^j_{\mu}W^k_{\nu}\,,
\end{equation}
\begin{equation}
F_{\mu\nu}(W^0) = \partial_{\mu}W^0_{\nu} -
\partial_{\nu}W^{0}_{\mu}\,,
\end{equation}
\begin{equation}
F^{a}_{\mu\nu}(G) = \partial_{\mu}G^a_{\nu} - \partial_{\nu}G^a_{\mu}
+g_sf^{abc}G^b_{\mu}G^c_{\nu}\,,
\end{equation}
where $W^i_{\mu}$, $W^0_{\mu}$ are the $SU_L(2) \otimes U(1)$ gauge
fields, $G^a_{\mu}$ are the gluon fields and $\epsilon^{ijk}$,
$f^{abc}$ are the structure constants of the $SU(2)$ and $SU(3)$
gauge groups. The lagrangian $L_{HYM}$ describes the Higgs doublet
interaction with $SU_L(2)\otimes U(1)$ gauge fields
\begin{equation}
L_{HYM} = (D_{L\mu}H)^{+}(D^{\mu}_LH)\,,
\end{equation}
where covariant derivatives are given by
\begin{equation}
D_{L\mu} = \partial_{\mu} -ig_1\frac{Y}{2}W^0_{\mu} - ig_2
\frac{\sigma^{i}}{2}W^i_{\mu}\,,
\end{equation}
\begin{equation}
D_{R\mu} = \partial_{\mu} -ig_1\frac{Y}{2}W^0_{\mu}\,,
\end{equation}
\begin{equation}
D^q_{L\mu} = \partial_{\mu} - ig_1\frac{Y}{2}W^0_{\mu} - ig_2
\frac{\sigma^{i}}{2}W^{i}_{\mu} - ig_st^aG^a_{\mu}\,,
\end{equation}
\begin{equation}
D^q_{R\mu} = \partial_{\mu} - ig_1\frac{Y}{2}W^0_{\mu} -
ig_st^aG^{a}_{\mu}\,.
\end{equation}
Here $g_1$ is the $U(1)$ gauge coupling constant, $Y$ is the hypercharge
determined by the relation $Q = \frac{\sigma_{3}}{2} + \frac{Y}{2}$,
$\sigma^{i}$ are Pauli matrices, $t^{a}$ are $SU(3)$ matrices in the
fundamental representation,
$H = \left( \begin{array}{cc}
H_1\\
H_2
\end{array}\right)$ 
is the Higgs $SU(2)$ doublet with $Y = 1$. The
lagrangian $L_{SH}$ describing Higgs doublet self-interaction has the
form
\begin{equation}
L_{SH} = -V_0(H) = M^2H^{+}H - \frac{\lambda}{2}(H^{+}H)^2\,,
\end{equation}
where $H^{+}H = \sum_{i}H^{*}_iH_i$ and $\lambda$ is the Higgs
self-coupling constant. Lagrangian $L_{f}$ describes the interaction
of fermions with gauge fields. Fermions constitute only doublets and
singlets in $SU_L(2) \otimes U(1)$
\begin{equation}
R_1 = e_R,\;R_2 =\mu_{R},\;R_{3} = \tau_{R} \,,
\end{equation}
\begin{equation}
L_1 = {\nu \choose e}_L \; L_2 ={\nu^{'} \choose \mu}_L \;
L_3 = {\nu^{''} \choose \tau}_L\,
\end{equation}
\begin{equation}
R_{qIu} = (q_{Iu})_R, \;\; (q_{1u} = u, \; q_{2u} = c, \; q_{3u} = t)\,,
\end{equation}
\begin{equation}
R_{qid} =(q_{id})_R, \;\; (q_{1d} = d, \; q_{2d} = s, \; q_{3d} =b)\,,
\end{equation}
\begin{equation}
L_{qI} = {q_{Iu} \choose V^{-1}_{Ii}q_{id}}_L \,,
\end{equation}
where L and R denote left- and right-handed components of the
spinors respectively,
\begin{equation}
\psi_{R,L} = \frac{1 \pm \gamma_{5}}{2} \psi
\end{equation}
and $V_{iI}$ is the Kobayashi-Maskawa matrix. The neutrinos are
assumed to be left-handed and massless. The Lagrangian $L_{f}$ has
the form
\begin{equation}
L_{f} =\sum_{k = 1}^{3}[ i\bar{L}_{k}\hat{D}_{L}L_{k} +
i\bar{R}_{k}\hat{D}_{R}R_{k} + i\bar{L}_{qk}\hat{D}^q_LL_{qk}
+ i\bar{R}_{qku}\hat{D}^q_RR_{qku} +
i\bar{R}_{qkd}\hat{D}^q_RR_{qkd}] \,,
\end{equation}
where $\hat{D}_L = \gamma^{\mu}D_{L\mu}$, $\hat{D}_{R} =
\gamma^{\mu}D_{R\mu}$, $\hat{D}^q_L = \gamma^{\mu}D^q_{L\mu}$,
$\hat{D}^q_R = \gamma^{\mu}D^q_{R\mu}$.
The Lagrangian $L_{Yuk}$ generates fermion mass terms. Supposing the
neutrinos to be massless,  the Yukawa interaction of the
fermions with Higgs doublet has  the form
\begin{equation}
L_{Yuk} = -\sum_{k=1}^{3}[h_{lk}\bar{L}_kHR_k + h_{dk}\bar{L}_{dk}H
R_{dk} + h_{uk}\bar{L}_{uk}(i\sigma^{2}H^{*})R_{uk}] \,+ h.c.\,.
\end{equation}
The potential term $V_0(H)= -M^2H^{+}H  +
\frac{\lambda}{2}(H^{+}H)^2$
for $M^2 > 0$  gives rise to the
spontaneous  symmetry breaking. The doublet $H$ acquires the nonzero
vacuum expectation value
\begin{equation}
<H> = \left( \begin{array}{cc}
0\\
\frac{v}{\sqrt{2}}
\end{array} \right)\,,
\end{equation}
where $v = 246 $ Gev. In the unitare gauge unphysical Goldstone
massless fields are absent and the Higgs doublet scalar field depends
on the single physical scalar field $h(x)$ (Higgs field):
\begin{equation}
H(x) = \left( \begin{array}{cc}
0\\
\frac{v}{\sqrt{2}} + \frac{h(x)}{\sqrt{2}}
\end{array} \right)\,.
\end{equation}
Due to spontaneous gauge symmetry breaking gauge fields except photon
field acquire masses. Diagonalization of mass matrix gives
\begin{equation}
W_{\mu}^{\pm} = \frac{1}{\sqrt{2}}(W^1_{\mu} \mp W^2_{mp}), \; M_W =
\frac{1}{2}g_2v \,,
\end{equation}
\begin{equation}
Z_{\mu} = \frac{1}{\sqrt{g^2_2 + g^2_1}}(g_2W^3_{\mu} -g_1W^0_{\mu}),
\; M_Z = \frac{1}{2}\sqrt{g^2_2 +g^2_1}v\,,
\end{equation}
\begin{equation}
A_{\mu} = \frac{1}{\sqrt{q^2_2 + g^2_1}}(g_1W^3_{\mu} + g_2W^0_{\mu}),
\; M_A = 0 \,,
\end{equation}
where $W^{\pm}_{\mu}$, $Z_{\mu}$ are charged and
neutral electroweak bosons,
$A_{\mu}$ is photon. It is convenient to introduce rotation angle
$\theta_{W}$ between $(W^3,W^0)$ and $(Z,A)$ which is called Weinberg
angle
\begin{equation}
\sin{\theta_{W}} \equiv \frac{g_1}{\sqrt{g^2_1 + g^2_2}} \,.
\end{equation}
Experimentally $\sin^{2}{\theta_{W}} \approx 0.23$ \cite{11}.
The formula for the electric charge $e$ has the form
\begin{equation}
e = \frac{g_2g_1}{\sqrt{g^2_2 + g^2_1}}\,.
\end{equation}
At the tree level the Higgs boson mass is determined by the
formula
\begin{equation}
m_h = \sqrt{2} M = \sqrt{\lambda} v \,.
\end{equation}
The lagrangian $L_{Yuk}$ is responsible for the fermion masses
generation. In the unitare gauge the lagrangian $L_{HYM}$ takes the
form
\begin{equation}
L_{HYM} = \frac{1}{2}\partial^{\mu}h \partial_{\mu}h +
M^2_W(1 + \frac{h}{v})^2 W^{+}_{\mu}W^{\mu} + \frac{1}{2}
M^2_Z(1 + \frac{h}{v})^2Z^{\mu}Z_{\mu} \,.
\end{equation}
The Yukawa lagrangian in the unitare gauge can be written
in the form
\begin{equation}
L_{Yuk} = -\sum_{i} m_{\psi_{i}}(1 + \frac{h}{v})
 \bar{\psi}_{i} \psi_{i} \,.
\end{equation}

At present LEP1 lower bound on the Higgs boson mass is \cite{12}
\begin{equation}
m_h > 63.5\,Gev \,.
\end{equation}
In standard  Weinberg-Salam model there are several theoretical
bounds on the Higgs boson mass :

(i) Tree level unitarity requirement leads to $m_h \leq 1$ Tev
\cite{13}.

(ii) The requirement of the absence of the Landau pole singularity
for the effective Higgs self-coupling constant for energies up to
$10^{14}$  Gev gives $m_h \leq 200$ Gev for $m_t \leq 200$ Gev
\cite{14}.

(iii) The vacuum stability requirement leads to the lower
bound on the Higgs boson mass which depends on the top quark
mass \cite{15}.

The renormalization group equations for the effective coupling
constants in neglection of all Yukawa coupling constants except
top-quark Yukawa coupling constant
in one-loop approximation read
\begin{equation}
\frac{d\bar{g}_3}{dt} = -7\bar{g}^3_3\,,
\end{equation}
\begin{equation}
\frac{d\bar{g}_2}{dt} = -(\frac{19}{6})\bar{g}^3_2\,,
\end{equation}
\begin{equation}
\frac{d\bar{g}_1}{dt} = (\frac{41}{6})\bar{g}^3_1\,,
\end{equation}
\begin{equation}
\frac{d\bar{h}_t}{dt} = (\frac{9\bar{h}^2_t}{2} -8\bar{g}^2_3 -
\frac{9\bar{g}^2_2}{4} -\frac{17\bar{g}^2_1}{12})\bar{h}_t\,,
\end{equation}
\begin{equation}
\frac{d\bar{\lambda}}{dt} = 12(\bar{\lambda}^2 + (\bar{h}^2_t -
\frac{\bar{g}^2_1}{4} - \frac{3\bar{g}^2_2}{4})\lambda -
\bar{h}^4_t
+ \frac{\bar{g}^4_1}{16} + \frac{\bar{g}^2_1\bar{g}^2_2}{8} +
\frac{3\bar{g}^4_2}{16})\,,
\end{equation}
\begin{equation}
t = (\frac{1}{16\pi^2})\ln{(\mu/m_Z)}\,.
\end{equation}
Here $\bar{g}_3$, $\bar{g}_2$ and $\bar{g}_1$ are the $SU(3)$,
$SU_L(2)$ and $U(1)$ gauge couplings, respectively, and $\bar{h}_t$ is
the top quark Yukawa coupling constant. In our estimates we
took $m_t^{pole} = 175$ Gev, $\bar{\alpha}_3(m_Z) = 0.118$,
$\bar{\alpha}_{em}^{-1}(m_Z) = 127.9$, $\sin^{2}{\theta_W}(m_Z) =
0.2337$, $\alpha_i \equiv \frac{g^2_i}{4\pi}$. From the requirement
of the absence of Landau pole singularity for the Higgs self-coupling
constant $\lambda$ for the scales  up to $\Lambda = (10^{3}; 10^{4};
10^{6}; 10^{8}; 10^{10}; 10^{12}; 10^{14})$ Gev
(to be precise we require that at the scale $\Lambda$ the Higgs
self-coupling constant is $\frac{\bar{\lambda}^{2}(\Lambda)}{4\pi} \leq
1$) we have found the upper bound on the Higgs boson mass $m_h \leq
(400; 300; 240; 200; 180; 170; 160)$ Gev, respectively. Roughly
speaking, the vacuum stability bound comes from the requrement that
the Higgs self-interaction coupling is nonegative
$\bar{\lambda}(\mu) \geq 0$ for the scales $\mu \leq M_s$. Here
$M_s$ is the scale up to which the standard model is applicable.
Suppose that at scales up to $M_s$ standard model works and at
scales $M \geq M_s$ we have some supersymmetric extension of the
standard model. It should be noted that the most popular at present
the minimal supersymmetric standard model (MSSM) predicts that
the effective Higgs self-coupling constant for the standard model
at the scale  of supersymmetry breaking $M_s$ has to obey the
inequality
\begin{equation}
0 \leq \bar{\lambda}(M_s) = (\bar{g}^2_1(M_s) + \bar{g}^2_2(M_s))
(\cos(2\varphi))^2/4 \leq (\bar{g}^2_1(M_s) + \bar{g}^2_2(M_s))/4 \,.
\end{equation}
So the assumption that standard Weinberg-Salam model originates
from its supersymmetric extension with the supersymmetry broken
at scale $M_s$ allows us to obtain non-trivial information about
the low energy effective Higgs self-coupling constant in the
effective potential $V = - M^2H^{+}H  +
\frac{\lambda}{2}(H^{+}H)^2$ and hence to obtain nontrivial
information about the Higgs boson mass. It should be noted that
in nonminimal supersymmetric electroweak models, say in the model
with additional gauge  singlet $\sigma$, we have due to the
$k\sigma H_1 i \tau_{2}H_2$ term in the superpotential an additional
term $k^2|H_1i\tau_{2}H_{2}|^2$ in the potential and as a consequence
our boundary condition for the Higgs self-coupling constant has to be
modified, namely
\begin{equation}
\bar{\lambda}(M_s) = \frac{1}{4}(\bar{g}^2_1(M_s) + \bar{g}^2_2(M_s))
\cos^2(2\varphi)
+ \frac{1}{2}\bar{k}^2(M_s)\sin^{2}(2\varphi) \geq 0 \,.
\end{equation}
The boundary condition  (72) depends on unknown coupling constant
$\bar{k}^2(M_s)$. However it is very important to stress that for all
nonminimal supersymmetric models broken to standard Weinberg-Salam
model at scale $M_s$ the effective Higgs self-coupling constant
$\bar{\lambda}(M_s)$  is  non-negative which is a direct
consequence of the non-negativity of the effective potential
in supersymmetric models. Therefore the vacuum stability
requirement results naturally \cite{16} if supersymmetry is broken at
some high scale $M_s$ and at lower scales standard Weinberg-Salam
model is an effective theory. For MSSM with boundary condition
(71) for the Higgs self-coupling constant $\bar{\lambda}(M_s)$
we have integrated numerically renormalization group equations
in two-loop approximation. Also we took into account the
one loop correction to the Higgs boson mass
 (running Higgs boson mass $\bar{m}_h(\mu) =
\sqrt{\bar{\lambda}(\mu)}v$ does not coincide with pole
Higgs boson mass). Our results for the Higgs boson mass
for different values of $M_s$ and  $m_t^{pole}$ are presented in
table 1. Here $k = 0$ corresponds to the boundary condition
$\bar{\lambda}(M_s) = 0$ and $k = 1$ corresponds to
the boundary condition $\bar{\lambda}(M_s) = \frac{1}{4}
(\bar{g}^2_1 + \bar{g}^2_2)$.

\newpage

Table 1. The dependence of the Higgs boson mass on the values of
$M_s$, $m_t^{pole}$ and $k =0,1$. Everything except k is in Gev.

\begin{center}
\begin{tabular}{|l| |l| |l| |l| |l| |l| |l| |l| |l| |l| |l|}
\hline
$m_t^{pole}$ & 165 & 165 & 170 & 170 &175 & 175 & 180 & 180 & 185 & 185 \\
\hline
 &k=0 &k=1& k=0 & k=1 &k=0& k=1 &k=0& k=1& k=0& k=1\\
\hline
$M_s=10^{3}$&69&111&74&114&78&117&83&120&88&123\\
\hline
$M_s=10^{3.5}$&81&117&86&120&92&124&98&128&104&132\\
\hline
$M_s=10^{4}$&89&121&95&125&101&130&108&134&114&139\\
\hline
$M_s=10^{6}$&105&129&113&135&121&141&129&147&137&153\\
\hline
$M_s=10^{8}$&112&132&120&138&129&147&138&152&146&159\\
\hline
$M_s=10^{10}$&115&133&124&140&133&147&142&154&151&161\\
\hline
$M_s=10^{12}$&117&134&126&141&136&147&145&154&154&161\\
\hline
$M_s=10^{14}$&118&134&127&141&132&148&147&156&156&164\\
\hline
$M_s=10^{16}$&118&134&128&141&138&148&148&156&158&164\\
\hline
\end{tabular}
\end{center}
\newpage

The tree-level Higgs boson couplings to gauge bosons and fermions
can be deduced from the lagrangians (62, 63). Of these, the
$hW^+W^-$, $hZZ$ and $h\bar{\psi} \psi$ are the most important for
the phenomenology. The partial decay width into fermion-antifermion
pair is \cite{10}
\begin{equation}
\Gamma(h \rightarrow \psi \bar{\psi}) = \frac{G_Fm^2_{\psi}m_hN_c}
{4\pi \sqrt{2}}(1 -\frac{4m^2_{\psi}}{m^2_h})^{\frac{3}{2}}\,,
\end{equation}
where $N_c$ is the number of fermion colours. For $m_{h} \leq 2m_W$
Higgs boson decays mainly with ($\approx$ 90 percent) probability
into b quark-antiquark pair and with $\approx$ 10 percent probability
into $\tau$ lepton-antilepton pair. An account of higher order QCD
corrections can be effectively taken into account in the formula
(73) for the Higgs boson decay into b quark-antiquark pair by the
replacement of pole b-quark mass in formula (73) by the effective
b-quark mass $\bar{m}_b(m_h)$.
Higgs boson with $m_h \geq 2M_{W}$ will decay into pairs of
gauge bosons with the partial widths
\begin{equation}
\Gamma(h \rightarrow W^{+}W^{-}) = \frac{G_Fm^3_h}{32\pi\sqrt{2}}
(4 -4a_w +3a^2_w)(1-a_w)^{\frac{1}{2}} \,,
\end{equation}
\begin{equation}
\Gamma(h \rightarrow Z^0 Z^0) = \frac{G_Fm^3_h}{64\pi\sqrt{2}}
(4 - 4a_Z + 3a^2_Z)(1 -a_Z) ^{\frac{1}{2}} \,,
\end{equation}
where $a_W = \frac{4M^2_W}{m^2_h}$ and $a_Z=\frac{4M^2_Z}{m^2_h}$.
In the heavy Higgs mass regime $(2m_Z \leq m_h \leq 800$ Gev), the
Higgs boson decays dominantly into gauge bosons. For example,
for $m_h \gg 2m_Z$ one can find that
\begin{equation}
\Gamma(h \rightarrow W^{+}W^{-}) = 2\Gamma(h \rightarrow ZZ)
\simeq \frac{G_Fm^2_h}{8\pi \sqrt{2}}\,.
\end{equation}
The $m^3_h$ behaviour is a consequence of the longitudinal
polarization states of the $W$ and $Z$. As $m_h$ gets large, so does
the coupling of $h$ to the Goldstone bosons which have been eaten by
the $W$ and $Z$. However, the Higgs boson decay width to a pair of
heavy quarks growth only linearly in the Higgs boson mass. Thus, for
the Higgs masses sufficiently above $2m_Z$, the total Higgs boson
width is well approximated by ignoring the Higgs decay to $t\bar{t}$
and including only the two gauge boson modes. For heavy Higgs
boson mass one can find that
\begin{equation}
\Gamma_{total}(H) \simeq 0.48\,Tev(\frac{m_h}{1\,Tev})^3 \,.
\end{equation}
It should be noted that there are a number of Higgs couplings
which are absent at tree level but appear at one-loop level.
Among them the couplings of the Higgs boson  to two
gluons and two photons are extremely important for the Higgs
boson searches at supercolliders. One-loop induced Higgs
coupling to two gluons is due to t-quark exchange in the loop
\cite{17} and it leads to an effective Lagrangian
\begin{equation}
L^{eff}_{hgg} = \frac{g_2\alpha_s}{24\pi m_W}hG^a_{\mu\nu}
G^{a\mu\nu} \,.
\end{equation}
Using the effective Lagrangian (78) one can find that
\begin{equation}
\Gamma(h \rightarrow gg) = \frac{g^2_2\alpha^{2}_{s}m^3_h}{288\pi^{3}
m^2_W}  \,.
\end{equation}
Also very important is  the one-loop induced Higgs boson coupling
to two photons due to W- and t-quark exchanges in the loop. The
corresponding expression for the decay width of the Higgs boson
into two photons is contained in  \cite{18}.

Consider now the Higgs boson search at supercolliders. For
completness let us start with LEP2. At LEP2 with the total
energy $\sqrt{s} =192$ Gev the dominant Higgs production
process is $e^+e^- \rightarrow hZ$. The corresponding cross section
at tree level is given by \cite{19}
\begin{equation}
\sigma(e^+e^- \rightarrow hZ) =
\frac{\pi \alpha^{2}\lambda^{1/2}
(\lambda + 12sM^2_Z)
[1 + (1 - 4\sin^{2}{{\theta}_W})^2]}
{192s^2\sin^4{{\theta}_W} \cos^4{{\theta}_W}(s -M^2_Z)^2}\,,
\end{equation}
where $\lambda \equiv (s - m^2_h -M^2_Z)^2 - 4m^2_hM^2_Z$. One can
see that for a fixed value of $m_h$, the cross section is maximal
for $\sqrt{s} \approx m_Z +\sqrt{2}m_h$. For $L_t =500$ $pb^{-1}$
one expects to observe the Higgs bosons using the signature
$Z \rightarrow \nu \bar{\nu}$, $h \rightarrow b \bar{b}$ for masses
up to $M_Z$. Note that the region $m_h \simeq M_Z$ is particularly
troublesome due to the $e^+e^- \rightarrow ZZ$ background. Both $ZZ$
and $hZ$ lead to four-fermion states. However, these can be separated
by making use of the fact that $Br(h \rightarrow b \bar{b})
\approx 90$ percent
as compared with $Br(Z \rightarrow b\bar{b}) \approx 20$ percent.
Given sufficient energy and luminosity, and a vertex detector
that can tag b-quark jets with high efficiency, it would be possible
to discover a Higgs boson if it is degenerate in mass with $Z$-boson.

At LHC the dominant mechanism for the Higgs boson production is
gluon-gluon fusion. Also $WW $ fusion becomes important for heavy
Higgs boson.
The cross section for the Higgs boson  production due to  gluon-gluon
fusion has the form
\begin{equation}
\frac{d\sigma}{dy}(AB \rightarrow h + X) =
\frac{\pi^{2}\Gamma(h \rightarrow gg)}{8m_hs}G_{A}(x_a,m^2_h)
G_{B}(x_b,m^2_h) \,,
\end{equation}
where $G_A$ and $G_B$ are the gluon distributions functions in
hadrons $A$ and $B$ respectively, and
\begin{equation}
x_a = \frac{m_he^y}{\sqrt{s}}, \; x_b = \frac{m_he^{-y}}{\sqrt{s}}\,,
\end{equation}
\begin{equation}
y = \frac{1}{2} \ln{(\frac{E_h + p_{3h}}{E_h -p_{3h}})}\,.
\end{equation}
The rapidity $y$ is defined in terms of the Higgs boson energy
and longitudinal momentum, defined in the laboratory frame.

\paragraph{3.2.3 Search for standard Higgs boson at CMS}

\subparagraph{The search for $h \rightarrow \gamma \gamma$.}

One of the most important reactions for the search for Higgs boson at
LHC is
\begin{equation}
pp \rightarrow (h \rightarrow \gamma\gamma) +...\,,
\end{equation}
which is the most promising one for the search for Higgs boson in the
most interesting region 90 Gev$ \le m_{h} \le $140 Gev.

The key features that enable CMS detector to obtain clear two-photon
mass peaks, significantly above background throughout the
intermediate mass range, are:

i. An electromagnetic calorimeter with an excellent energy resolution
(this requires calibration to high precision, which in turn requires
a good inner tracking system).

ii. A large acceptance (the precision electromagnetic calorimetry
extends to $|\eta| = 2.5$), adequate neutral pion rejection and (at
high luminosity) a good measurement of photon direction. This
requires fine lateral segmentation and a preshower detector.

iii. Use of powerful inner tracking system for isolation cuts.

The cross section times branching has been estimated to be
$\sigma Br(h \rightarrow \gamma \gamma ) = 76 fb(68fb)$
for $m_h =110(130)$ Gev, the uncertainty in
the cross section calculation is
(30 - 50) percent. The imposition of cuts ($|\eta| \le 2.5$,
$p_{T}^{\gamma_1} \geq 40$ Gev, $p_{T}^{\gamma_2} \geq 25$ Gev)
allow to decrease the background in a reasonable magnitude. The jet
background is reduced by imposing an isolation cut, which also
reduces the bremsstrahlung background. Photon is defined to be
isolated if there is no charged track or electromagnetic shower
with a momentum greater than 2.5 Gev within a region $\Delta R
\leq 0.3$ around it. The photons from the decay of $\pi^{0}$ of the
relevant transverse momenta are separated in the calorimeter by a
lateral distance of the order of 1 cm. An efficiency of 64 percent was
assumed for reconstruction of each photon (i.e. 41 percent per
event). The crystal calorimeter was assumed to have an energy
resolution $\Delta E/E = 0.02/\sqrt{E} \oplus 0.005 \oplus 0.2/E$ in
the barrel and $\Delta E/E = 0.05/\sqrt{E} \oplus 0.005 \oplus 0.2/E$
in the endcap, where there is a preshower detector. At high
luminosity, a barrel pre-shower detector covers $|\eta| < 1.1$,
resulting in  a resolution $ \Delta E/E = 0.05/\sqrt{E} \oplus 0.005
\oplus 0.2/E$ and an ability to measure the photon direction with
resolution $\Delta \alpha = 40$ $mrad/\sqrt{E}$ in this region.

The background to the $h \rightarrow \gamma \gamma$ may  be divided
into 3 categories:

1. prompt diphoton production from quark annihilation and gluon
 fusion diagrams - irreducible background,

2. prompt diphoton production from bremsstrahlung from the outgoing
 quark line in the QCD Compton diagram,

 3. background from jets, where an electromagnetic energy deposit
 originates from the decay of neutral hadrons in a jet from 1 jet +
 1 prompt photon.

 The signal  significance $\sigma =
 \frac{N_S}{\sqrt{N_B}}$ is estimated to be $7\sigma(9\sigma)$ for
 $m_h = 110(130) $Gev and for low luminosity $L_{low,t} = 3\cdot
 10^{4}pb^{-1}$ and $10\sigma(13\sigma)$ for $m_h = 110(130)$ Gev and
 for high luminosity $L_{high,t} = 10^{5}pb^{-1}$. The general
 conclusion is that at $5\sigma$ level it would be possible to discover
 Higgs boson for 95 Gev$ \le m_{h} \le 145$ Gev
at low luminosity and at high
 luminosity the corresponding Higgs boson mass discovery interval is
 85 Gev$ \le m_{h} \le 150$ Gev.

\subparagraph{ Search for $h \rightarrow \gamma\gamma$ in association
with high $E_{T}$ jets.}

The possibility of the search for $h
\rightarrow \gamma\gamma$ with large $E_{T}$ jet allows to improve
 $signal/background$ ratio. There are several sources of such Higgs +
 jet events. One is the next to leading order corrections to $gg
 \rightarrow h$ with hard gluons.  Others are the associated
 production of $\bar{t}th, Wh, Zh$ and the WW and Zh fusion
 mechanisms.

The cuts that provide optimal sensitivity are \cite{1}:

i. Two isolated photons are required, with $p_t^{\gamma_{1}}
\geq 40$ Gev and $p_{t}^{\gamma_{2}} \geq 60$ Gev, $|\eta| \leq 2.5$ and
$p_t^{\gamma \gamma} \geq 50$ Gev.

ii. Number of jets $\geq 2$, $E_t^{jet} \geq 40$ Gev for the
central jets ( $|\eta| \leq 2.4$ ) and $E^{jet} \geq 800$ Gev
for the forward ones ($2.4 \leq |\eta| \leq 4.6$).

iii. Photons are isolated with no charged or neutral particles with
$p_t \geq 2$ Gev within a cone $\Delta R \leq 0.3$ around each
photon's direction.

iiii. $\gamma$-jet isolation $\Delta R(\gamma ,jet) > 1.5$ (to
suppress the bremsstrahlung contribution).

The calculations give encouraging results, namely for
 $L_{high,t} = 1.6\cdot10^{5}pb^{-1}$ it would be possible to
 discover the Higgs boson for 70 Gev$ \le m_h \le $150 Gev with $\ge
 7\sigma$ signal significance. Note that  the background is not only
much smaller in magnitude than in the inclusive $h \rightarrow
\gamma \gamma $ search, but it is also peaked at higher masses, away
from the most difficult region $m(\gamma \gamma) \leq$ 90 Gev.

\subparagraph{$h \rightarrow \gamma \gamma$ in
 associated Wh and $\bar{t}th$ production.}

The $Wh \rightarrow
l\gamma\gamma + X $ and $\bar{t}th \rightarrow l\gamma\gamma + X$
final states are other promising signature for the Higgs boson
search. The production cross section is smaller than the inclusive
$h \rightarrow \gamma \gamma $ by a factor $\approx 30$. However
the isolated hard lepton from the W and t decay allows to obtain a
strong background reduction and to indicate the primary vertex at
any luminosity. The main backgrounds are: $W\gamma \gamma$,
 $W \gamma + jet$, $W + 2jets$, $t\bar{t}\gamma \gamma$, $t\bar{t}
 \gamma$, $t\bar{t}$, $b\bar{b}$, $c\bar{c}$ and $t\bar{b}$, with
 a real lepton in the final state and one or two jets faking photons
 due to hard bremsstrahlung.
For $80 Gev \le m_{h} \le 120 Gev$ and $L_{high,t} =
1.6\cdot10^{5}pb^{-1}$ the signal significance is better than
 $6\sigma$.

\subparagraph{$h \rightarrow ZZ^{*}(ZZ) \rightarrow 4$ leptons.}

The search for the standard Higgs boson using the reaction
$h \rightarrow ZZ^{*}(ZZ) \rightarrow 4$ leptons is possible for a
broad mass range 130 Gev$ \le m_{h} \le $800 Gev.
Below  $2M_Z$    the event rate is small and the background
reduction more difficult, as one of the $Z$s is off mass shell. In
this mass region the width of the Higgs boson is small $\Gamma_{h}
< 1$ Gev. The significance of the signal is proportional to the
mass resolution, so the lepton momentum resolution is of decisive
importance. The geometrical and kinematic acceptance for leptons is
also very important in these channels \cite{1}
in the $m_h <2M_Z$ mass region , the main backgrounds are from
$t\bar{t}$, $Zb\bar{b}$ and $ZZ^{*}$. The $ZZ^{*}$ background is
irreducible and peaks sharply near the $ZZ$ threshold. The
$Zb\bar{b}$ background cannot be reduced by a $Z$ mass cut, but it
can be suppressed by lepton isolation. The $t\bar{t}$ background
can be reduced by a $Z$ mass cut and by isolation cuts. The standard
event cuts were choosen the following \cite{1}:
one electron with $p_t > 20$ Gev; one with $p_t > 15$ Gev, and the
remaining two electrons with $p_t >10$ Gev, all within $|\eta| <
2.5$. For muons, the corresponding $p_t$ cuts are 20, 10 and 5 Gev in
the rapidity range $|\eta| <2.4$. For $m_h = 130$ Gev the overall
(kinematic and geometrical) acceptance for the four-electron channel
is 22 percent and for the four-muon channel 42 percent. For $m_h =
170$ Gev these acceptances increase to 38 percent and 48 percent
respectively. To select $h \rightarrow ZZ^{*}$ events and
suppress the large $t\bar{t}$ background, one of the $e^{+}e^{-}$
or $\mu^{+}\mu^{-}$ pairs was assumed to be within $\pm 2\sigma_{Z}$
of the $Z$ mass. There is a fraction of events where both $Z$s are
off-shell. This effect results in a 24 percent loss for $m_h = 130$
Gev, decreasing to 12 percent for $m_h = 170$ Gev. The $M_Z$ cut
reduce $t\bar{t}$ background by a factor 11 in the $Z \rightarrow
\mu^{+}\mu^{-}$ channel and by a factor of 5 in the $Z \rightarrow
e^{+}e^{-}$ channel. For  two softer leptons, $M(ll) > 12$ Gev
is also required.

For the region 130 Gev$ \le m_{h} \le $180 Gev it would be
possible for $L_{high,t} = 10^{5}pb^{-1}$ to discover the Higgs boson
with $\ge 5\sigma$ signal significance except narrow mass region
around 170 Gev. For $m_{h} \ge $ 180 Gev the $h \rightarrow ZZ
 \rightarrow 4l^{\pm}$ channel is sensitive for low luminosity
 $10^{4}pb^{-1}$ from $2m_{Z}$ to 400 Gev. The main background here
 is non-resonant $ZZ$ production. The conclusion is that even at low
 luminosity $L_{low,t} = 10^{4}pb^{-1}$ it would be possible to
 discover Higgs boson for $2m_{Z} \le m_{h} \le $ 400 Gev.

  \subparagraph{The use of channels $h \rightarrow ll\nu\nu$,
$h \rightarrow WW \rightarrow l \nu jj$ and $h \rightarrow ZZ
\rightarrow ll jj$.}

The channel $h \rightarrow l l \nu \nu$ has a six times larger
branching than $h \rightarrow 4l^{\pm}$. The main background comes
from $ZZ$, $ZW$, $t\bar{t}$ and $Z + jets$.  The choosen cuts
are the following:

1. $E_t^{miss} \geq 100$ Gev.

2. Two leptons are required, with $p_t \geq 20$ Gev, $|\eta|
\leq 1.8$ and $p_t^{ll} \geq 60$ Gev.

3. $|M_Z - M_{ll}| \leq 6$ Gev.

4. No other isolated leptons with $p_t \geq 6$ Gev.

5. No central jets with $E_t \geq 150$ Gev.

6. No jets back-to-back with leptons (cosine of the angle between
the momentum of the lepton pair and sum of the momenta of the jets
is  $\geq -0.8$).

 7. $E_t^{miss}$ vector back-to-back with the lepton pair (cosine
 of the angle in the transverse plane between the two-lepton momentum
 and the missing transverse momentum $\leq 0.8$).

The conclusion is that using this mode it would be possible to
discover Higgs boson in the interval 400 Gev$
\le m_{h} \le $(800 - 900) Gev.

The channels $h \rightarrow WW \rightarrow l\nu jj$ and
$h \rightarrow ZZ \rightarrow ll jj$ are important in the
$m_h \approx 1$ Tev mass range, where the large $W,Z
\rightarrow q \bar{q}$ branching ratios must be used.
Also high lepton pairs  with $m_{ll}  \approx M_Z$ for
$h \rightarrow ZZ$ or a high $p_t$ lepton pair
plus large $E^{miss}_t$, for $h \rightarrow WW$
must be used. In addition,
two hard jets from the hadronic decays of $Z/W$ with
$m_{jj}  \approx M_{Z/W}$ are required. The backgrounds are:
$Z + jets$, $ZW$, $WW$, $t\bar{t}$, $WW$, $WZ$. For
$m_h \approx 1$ Tev  the Higgs boson is very broad ($\Gamma_{h}
\approx 0.5$ Tev and $WW/ZZ$ fusion mechanism represents about
50 percent of the total production cross section), therefore
forward-region signature is essential.
The appropriate cuts are the following:

i. $E_t^{miss} \geq 150$ Gev, $p^l_t \geq 150$ Gev, $p_t^W \geq 300$
Gev for $h \rightarrow WW$, or $p_t^l \geq 50$ Gev, $p_t^Z \geq 50$
Gev, $p_t^Z \geq 150$ Gev, $|m_{Z} - m_{ll}| \leq 10$ Gev for $h
\rightarrow ZZ$.

ii. $|m_{jj} - m _{W/Z}| \leq 15$ Gev for the central jet pair.

iii. $E_t^{jet} \geq 10$ Gev, $E^{jet} \geq 400$ Gev, $|\eta|
\geq 2.4$ for the two forward tagging jets.

The main conclusion is that the use of the reactions $h \rightarrow
WW \rightarrow l\nu jj$ and $h \rightarrow ZZ \rightarrow lljj$
allows to discover the heavy Higgs boson with a mass up to 1 Tev for
$L_{high,t} = 10^{5}pb^{-1}$.

\paragraph{3.2.4 B-physics in SM}

\subparagraph{General comments.}

The main task of B-physics investigation at LHC is the observation of
CP-violation.  For such a measurement high statistics are important
since the useful decay rates have branchings $(10^{-4} - 10^{-7})$.
At LHC $(10^{12} - 10^{13})$ $ b\bar{b}$ pairs will be produced per
year so the main problem here is to trigger and select the interesting
modes. The violation of CP symmetry is one of the most intriguing aspects
of high-energy physics. At present there is only a single measurement
of a CP-violation parameter: the measurement of  $\epsilon_{K}$ in
$K$ decays. The standard model description of CP violation is
very predictive - all CP-violating effects are related to the
phase $\delta$ of the Cabibbo-Kobayashi-Maskawa (CKM) matrix.
Note that in standard model also $\theta$-term
\begin{equation}
\theta \frac{\alpha_{s}}{4\pi}TrG_{\mu\nu}\tilde{G}^{\mu\nu}
\end{equation}
in QCD lagrangian violates CP-symmetry \cite{20}. $\theta$-term gives
nonzero contribution to electric dipole moment of the neutron
\cite{20}. The current experimental bound on the electric dipole
moment of the neutron $|d_n| \leq 1.1 \cdot 10^{-25}e \cdot cm$
implies that $|\theta| \leq 10^{-9}$ that corresponds to an
extreme fine-tuning of a parameter of the QCD Lagrangian.
Therefore there are good prospects that detailed investigation
of CP violation in B-decays will provide us important information
on the possible mechanisms of CP-violation.

Let us start with brief description of P, C and CP transformations
\cite{21}.
The parity transformation is a space-time transformation, under which
$t \rightarrow -t$ and $ \vec x \rightarrow -\vec x$. It changes the
sign of momenta, $\vec p \rightarrow -\vec p$, leaving spins
unchanged. For pseudoscalar mesons $P$ and $\tilde{P}$, the
parity transformation could be defined as
\begin{equation}
\hat{P}|P(\vec p)> = - |P(-\vec p)>\,,
\end{equation}
\begin{equation}
\hat{P}|\tilde{P}(\vec p)> = -|\tilde{P}(-\vec p)> \,.
\end{equation}
Charge conjugation relates particles and antiparticles, leaving all
space-time coordinates unchanged, i.e.
\begin{equation}
\hat{C}|P(\vec p)> = |\tilde{P}(\vec p)> \,,
\end{equation}
\begin{equation}
\hat{C} |\tilde{P}(\vec p)> = |P(\vec p)> \,.
\end{equation}
The combined $CP$ transformation acts on pseudoscalar mesons
in the following way:
\begin{equation}
\hat{C}\hat{P}|P(\vec p)> = -|\tilde{P}(-\vec p)> \,,
\end{equation}
\begin{equation}
\hat{C}\hat{P}|\tilde{P}(\vec p)> = -|P(-\vec p)> \,.
\end{equation}
For neutral $P^0$ and $\tilde{P}^0$ mesons one can construct the
$CP$ eigenstates

$|P^0_1> = \frac{1}{\sqrt{2}}(|P^0> - |\tilde{P}^0>)$,

$|P^0_2> = \frac{1}{\sqrt{2}}(|P^0> + |\tilde{P}^0>)$,

which obey

$\hat{C} \hat{P}|P^0_1> = |P^0_1>$,

$\hat{C} \hat{P}|P^0_2> =
-|P^0_2>$.

\subparagraph{CP violation in standard model.}

The standard model gauge group $SU_{C}(3) \otimes SU_{L}(2) \otimes
U(1)$ is spontaneously broken to $SU_{C}(3)\otimes U(1)$ due to
nonzero vacuum expectation value of the Higgs doublet. This gives the
masses to the W and Z bosons, as well as to the quarks and leptons.
The quark masses arise from the Yukawa couplings to the Higgs
doublet. The Yukawa interactions are written in terms of the weak
eigenstates $q^{'}$ of the quark fields. After the electroweak
symmetry breaking the quark fields are redefined so as to obtain
the mass terms in the canonical way. In the weak basis these
charged-current interactions have the form
\begin{equation}
L_{int} =
-\frac{g_2}{\sqrt{2}}(\bar{u}^{'}_{L},\bar{c}^{'}_{L},\bar{t}^{'}_{L})
\gamma^{\mu} \left( \begin{array}{ccc}
d^{'}_L\\
s^{'}_L\\
b^{'}_L
\end{array} \right)
W^{+}_{\mu} + h.c. \,.
\end{equation}
In terms of the mass eigenstates q the $L_{int}$ can be rewritten in
the form
\begin{equation}
L_{int} = - \frac{g_2}{\sqrt{2}}(\bar{u}_L, \bar{c}_L, \bar{t}_L)
\gamma^{\mu}V_{CKM}\left( \begin{array}{ccc}
d_L\\
s_L\\
b_L
\end{array} \right)
W^{+}_{\mu} + h.c. \,.
\end{equation}
The CKM mixing matrix $V_{CKM}$ is a unitary matrix in flavour space.
In the case of three generations, $V_{CKM}$ can be parametrized by
three Euler angles and six phases, five of which can be removed by
redefinition of the left-handed quark fields. So three angles
$\theta_{ij}$ and one observable complex phase $\delta$ remain in the
quark mixing matrix \cite{21}.

The "standard parametrization" of the CKM matrix is
\begin{equation}
V_{CKM} = \left( \begin{array}{ccc}
c_{12}c_{13} & s_{12}c_{13} & s_{13}e^{-i\delta}\\
-s_{12}c_{23} -c_{12}s_{23}s_{13}e^{i\delta} & c_{12}c_{23}
-s_{12}s_{23}s_{13}e^{i\delta} & s_{23}c_{13}\\
s_{12}s_{23} - c_{12}c_{23}s_{13}e^{i\delta} &   - c_{12}s_{23} -
s_{12}c_{23}s_{13}e^{i\delta} & c_{23}c_{13}
\end{array} \right) \,.
\end{equation}
Here $c_{ij} = \cos(\theta_{ij})$ and $s_{ij} = \sin(\theta_{ij})$.
The imaginary part of the mixing matrix is necessary to describe CP
violation in standard model. In general CP is violated in flavour
changing decays if there is no degeneracy of any two quark masses
and if the quantity $J_{CP} \neq  0$, where
\begin{equation}
J_{CP} = |Im(V_{ij}V_{kl}V^{*}_{il}V^{*}_{kj}|; \;i \neq k \; j \neq l
\, .
\end{equation}
It can be shown that all CP-violating amplitudes in the standard
model are proportional to $J_{CP}$.

For many applications it is more convenient to use an approximate
parametrization of the CKM matrix (Wolfenstein parametrization)
\cite{22}
\begin{equation}
V_{CKM} \simeq \left( \begin{array}{ccc}
1 - \frac{\lambda^2}{2} & \lambda & A\lambda^{3}(\rho - i\eta)\\
-\lambda & 1 - \frac{\lambda^{2}}{2} & A\lambda^{2}\\
A\lambda^{3}(1 - \rho - i\eta) & -A\lambda^{2} & 1
\end{array} \right)
\, + O(\lambda^{4}) \,.
\end{equation}
One can find that
\begin{equation}
J_{CP} \simeq A^{2} \eta \lambda^{6} \simeq 1.1
\cdot 10^{-4} \cdot A^{2} \eta \,.
\end{equation}
The two Wolfenstein parameters $\lambda$ and $A$ are experimentally
well determined \cite{23}
\begin{equation}
\lambda = |V_{us}| = 0.2205 \pm 0.0018 \,,
\end{equation}
\begin{equation}
A = |\frac{V_{cb}}{V^{2}_{us}}| = 0.80 \pm 0.04 \,.
\end{equation}
A simple way to take into account the implications of the
CKM matrix unitarity is provided by the so-called unitarity triangle
which uses the fact that the unitarity equation
\begin{equation}
V_{ij}V^{*}_{ik} = 0 \; (j \neq k)
\end{equation}
can be represented as the equation of a closed triangle in the
complex plane. There are six such triangles with the same area
\cite{24}
\begin{equation}
|A_{\Delta}| = \frac{1}{2}J_{CP} \,.
\end{equation}
Most useful from the phenomenological point of view is the
triangle relation
\begin{equation}
V_{ud}V^{*}_{ub} + V_{cd}V^{*}_{cb} + V_{td}V^{*}_{tb} = 0 \,,
\end{equation}
which determines the triangle with the angles
$\alpha$, $\beta$, $\gamma$ ($\alpha + \beta +
\gamma = \pi$). The present fit to data for triangle gives:
$\sin(2\alpha) \approx 0.7$, $\sin(2\beta) \approx 0.5$ ,
$\sin(\gamma) \approx 0.9$.

\subparagraph{Direct CP violation in weak decays.}

Consider two decay processes related to each other by a CP
transformation. Let $P$ and $\tilde{P}$ be CP-conjugated pseudoscalar
meson states, and $f$ and $\tilde{f}$ some CP-conjugated final
states:
\begin{equation}
\hat{C} \hat{P} |P> = e^{i\phi_{P}}|\tilde{P}>, \;
\hat{C} \hat{P} |f> = e^{i\phi_{f}}|\tilde{f}>
\end{equation}
The phases  $\phi_{P}$ and $\phi_{f}$ are arbitrary. The
CP-conjugated  amplitudes , $A$ and $\bar{A}$, can be written as
\begin{equation}
A = <f|H|P> = \sum_{i} A_{i} e^{i\delta_i} e^{i\phi_{i}} \,,
\end{equation}
\begin{equation}
\bar{A} = <\tilde{f}|H|\tilde{P}> = e^{i(\phi_{P} - \phi_{f})}
\sum_{i} A_{i}e^{i\delta_{i}}e^{-i\phi_{i}} \,,
\end{equation}
where $H$ is the effective Hamiltonian for weak decays, and $A_{i}$
are real partial amplitudes. The weak phases $\phi_{i}$ are
parameters of the CP-violating lagrangian. They appear in the
electroweak sector of the theory and enter $a$ and $\bar{A}$ with
opposite signs. The strong phases $\delta_{i}$ appear in scattering
amplitudes even if the Lagrangian is CP invariant and they enter $A$
and $\bar{A}$ with the same sign. It appears that the ratio
 \begin{equation}
 |\frac{\bar{A}}{A}| = |\frac{\sum_{i}A_ie^{i\delta_i}e^{i\phi_i}}
 {\sum_{i}A_{i}e^{i\delta_{i}}e^{-i\phi_i}}|
 \end{equation}
 is independent of phase conventions. The nonequality
 \begin{equation}
 |\frac{\bar{A}}{A}| \neq 1
 \end{equation}
 implies direct CP violation which comes from the interference of
 decay amplitudes leading to the same final states, that requires at
 least two partial amplitudes that differ in both the weak and strong
 phases.

Since the mixing is unavoidable in neutral meson decays the best way
to observe direct CP violation in the decays of charged mesons.
The CP asymmetry is defined in standard way:
\begin{equation}
 a_f = \frac{\Gamma(P^+ \rightarrow f) - \Gamma(P^- \rightarrow
 \bar{f})}{\Gamma(P^+ \rightarrow f) + \Gamma(P^- \rightarrow
 \bar{f})} = \frac{1 - |\frac{\bar{A}}{A}|^2}{1 +
 |\frac{\bar{A}}{A}|^2} \,.
 \end{equation}

 We have to consider non-leptonic decays since leptonic and
 semileptonic decays are usually dominated by a single diagram
 and complex phases cancel. Non-leptonic decays can receive both
 tree and penguin contributions \cite{25}. Penguin diagrams typically
 involve other weak phases than tree diagrams. To get large
 interference effects one needs partial amplitudes with similar
 amplitude. Therefore it is necessary to look for decays in which
 tree  contribution is suppressed by small CKM parameters that
 compensates the loop suppression of penguin diagrams.
 Another possibility is to consider tree forbidden decays, which can
 only proceed through penguin diagrams. Examples are
 $B^{\pm} \rightarrow K^{\pm}K$, $B^{\pm} \rightarrow K^{\pm}\phi
 $, $B^{\pm} \rightarrow K^{*\pm}\gamma$, $B^{\pm} \rightarrow
 \rho^{\pm}\gamma$.
At present there is no experimental evidence for direct CP violation.

\subparagraph{Indirect CP violation in the mixing of neutral mesons.}

The neutral mesons $P^0$ and $\tilde{P}^0$ can mix via common
decay channels:
\begin{equation}
 P^0 \leftrightarrow X \leftrightarrow \tilde{P}^0 \,.
\end{equation}
An arbitrary neutral meson state can be written as a superposition
 of the eigenstates $a|P^0> +b|\tilde{P}^0>$ which obey Schrodinger
 equation
 \begin{equation}
i\frac{d(a,b)}{dt} =\vec H(a,b) =(\vec M - \frac{i}{2}\vec
\Gamma)(a,b) \,,
\end{equation}
where $\vec M$ and $\vec \Gamma$ are Hermitian $2 \times 2$ matrices.
Since the Hamiltonian operator, $\vec H$, is not Hermitian, its
eigenstates
\begin{equation}
|P_{1,2}> = p|P^0> \pm q|\tilde{P}^0>; \; |p^2| + |q^2| = 1
\end{equation}
are not orthogonal, and the eigenvalues
\begin{equation}
\mu_j =M_j - \frac{i}{2}\Gamma_j; \; j=1,2
\end{equation}
are complex. The time evolution of the states $P_i$ is given by
\begin{equation}
|P_i(t)> = e^{-iM_it}e^{-\frac{1}{2}\Gamma_it}|P_i(0)> \,.
\end{equation}
One can show that the ratio
\begin{equation}
|\frac{q}{p}|^2 = |\frac{M_{12}^{*} -\frac{i}{2}\Gamma^{*}_{12}}
{M_{12} - \frac{i}{2}\Gamma_{12}}|^{2}
\end{equation}
is independent of phase conventions. The condition $|\frac{q}{p}|
\neq 1$ implies CP violation due to the fact that flavour eigenstates
are different from the CP eigenstates.
The following relations are valid:
\begin{equation}
(\Delta m)^2 -\frac{1}{4}(\Delta \Gamma)^2 =4|M_{12}|^2
-|\Gamma_{12}|^2 \,,
\end{equation}
\begin{equation}
\Delta m \cdot \Delta \Gamma = 4Re(M_{12}\Gamma^{*}_{12}) \,,
\end{equation}
\begin{equation}
\frac{q}{p} = -\frac{1}{2}\frac{\Delta m -\frac{i}{2}\Delta \Gamma}
{M_{12} - \frac{i}{2}\Gamma_{12}} = -2\frac{M_{12}^{*}
-\frac{i}{2}\Gamma^{*}_{12}}{\Delta m - \frac{i}{2}\Delta\Gamma} \,,
\end{equation}
where $\Delta m = m_2 - m_1$ and $\Delta\Gamma = \Gamma_2 -
\Gamma_1$. An alternative notation is define $\tilde{\epsilon}$ such
that
\begin{equation}
p =\frac{1 + \tilde{\epsilon}}{\sqrt{2(1 + |\tilde{\epsilon}|^2)}},\;
q = \frac{1-\tilde{\epsilon}}{\sqrt{2(1 +|\tilde{\epsilon}|^2)}},\;
\frac{q}{p} = \frac{1 -\tilde{\epsilon}}{1 + \tilde{\epsilon}} \,.
\end{equation}
Consider at first kaon system. Define the "short-lived" and
"long-lived" neutral kaon states as $K_{S} = K_1$ and $K_{L} = K_2$.
Experimentally,
\begin{equation}
 \Delta m_K = m_L -m_S = (3.510 \pm 0.018)\times 10^{-15} Gev \,,
\end{equation}
\begin{equation}
 \Delta\Gamma_{K} = \Gamma_L - \Gamma_S = -(7.361 \pm 0.010) \times
 10^{-15} Gev \, .
\end{equation}
 Define
\begin{equation}
 \frac{\Gamma^{*}_{12}}{M_{12}^{*}} = -|\frac{\Gamma_{12}}{M_{12}}|
 e^{i\varphi_{12}} \,.
\end{equation}
 From the last definition we find that
\begin{equation}
 |\frac{q}{p}|_{K} - 1 \simeq -2Re(\tilde{\epsilon}_K) \simeq
 -\varphi_{12} = O(10^{-3}) \,.
\end{equation}

For B-meson systems decay channels common to $B^0$ and $\tilde{B}^{0}$,
which are responsible for the difference $\Delta\Gamma_{B}$,
are known to have branching ratios of order $10^{-3}$ or less. Hence
 $|\frac{\Delta\Gamma_B}{\Gamma_B}| \leq O(10^{-2})$. The observed
 $B^0 -\tilde{B}^0$ mixing implies $\Delta m_B/\Gamma_B = 0.74 \pm
 0.04$  that means $|\Delta \Gamma_B| \ll \Delta m_B$.
 Therefore, the lifetime difference between the CP eigenstates is very
 small and it is possible to define these states as "light" and
 "heavy", $B_L = B_1$ and $B_H = B_2$. It follows that $|\Gamma_{12}|
 \ll |M_{12}|$, and to first order in $\Gamma_{12}/M_{12}$ one can
 find that
\begin{equation}
|\frac{q}{p}|_B - 1 \simeq
 -2Re(\tilde{\epsilon}_B) = O(10^{-2})\,.
\end{equation}

Since $B_L$ and $B_H$ have almost identical lifetimes, it is not possible
to produce selectively beams of $B_L$ and $B_H$ particles.  The time
evolution of an initially pure $B^0$ state is
\begin{equation}
 |B^0(t)> =
 e^{-im_Bt}e^{-\frac{1}{2}\Gamma_Bt}(\cos(\frac{1}{2}\Delta m_Bt)
 |B^0> + \frac{iq}{p}\sin(\frac{1}{2}\Delta m_Bt)|\tilde{B}^0>) \,,
 \end{equation}
 \begin{equation}
 |\tilde{B}^0(t)> =
 e^{-im_Bt}e^{-\frac{1}{2}\Gamma_Bt}(\cos(\frac{1}{2}\Delta
  m_Bt)|\tilde{B}^0> + \frac{ip}{q}\sin(\frac{1}{2}\Delta
 m_Bt)|B^0>) \,,
\end{equation}
 where $m_{H,L} =m_B \pm
 \frac{1}{2}\Delta m_B$ and $\Gamma_{H,L} \simeq \Gamma_{B}$.
 Defining the semileptonic asymmetry as
 \begin{equation}
 a^B_{SL} =
 \frac{\Gamma(\tilde{B}^0(t) \rightarrow l^+\bar{\nu}X) -
 \Gamma(B^0(t) \rightarrow l^-\nu X)}{\Gamma(\tilde{B}^0(t)
 \rightarrow l^+\bar{\nu} X) + \Gamma(B^0(t) \rightarrow l^-\nu X)} \,,
 \end{equation}
 one can find that
 \begin{equation}
 a^{B}_{SL} = \frac{1 -|q/p|^4}{1 + |q/p|^4} \simeq
 4Re(\tilde{\epsilon}_B) = O(10^{-2}) \,.
 \end{equation}
 At present we don't have experimental evidence for indirect CP
 violation in the B-meson system.

Consider decays of neutral mesons into CP eigenstates:
\begin{equation}
A = <f_{CP}|H|P^0>, \;\; A^{*} = <f_{CP}|H|\tilde{P}^0> \,.
\end{equation}
The condition $\lambda =\frac{q}{p}\cdot\frac{\tilde{A}}{A} \neq 1$
implies CP violation. The most theoretically favoured scenario is
when $\lambda = e^{i\phi_{\lambda}}$. In that case $\lambda$ is a
pure phase and hadronic uncertainties are cancelled. Define
the CP asymmetry as
\begin{equation}
a_{f_{CP}} = \frac{\Gamma(B^0(t) \rightarrow f_{CP}) -
\Gamma(\tilde{B}^0(t) \rightarrow f_{CP})}{\Gamma(B^0(t) \rightarrow
 f_{CP} ) + \Gamma(\tilde{B}^0(t) \rightarrow f_{CP})} \,.
\end{equation}
Taking into account that $|\frac{q}{p}|_B \simeq 1$ we find that
\begin{equation}
a_{f_{CP}} \simeq \frac{(1-|\lambda|^2)\cos(\Delta m_Bt) - 2Im \lambda
\sin(\Delta m_Bt)}{1 + |\lambda|^2} \,.
\end{equation}

Decays of neutral B mesons into CP eigenstates provide for
model-independent determination of CP-violating phase. In the
B-meson system we have
\begin{equation}
 (\frac{q}{p})_B \simeq - \frac{M_{12}^{*}}{|M_{12}|} =
\frac{(V_{tb}^{*}V_{td})^2}{|V^{*}_{tb}V_{td}|^2} =
 \frac{V^{*}_{tb}V_{td}}{V_{tb}V^{*}_{td}} = e^{-2i\beta} \,.
\end{equation}
To eliminate hadronic uncertainties it is necessary to choose decay
modes dominated by a single diagram. Examples of such decays are
$\tilde{B} \rightarrow \pi\pi$, $\tilde{B} \rightarrow D\tilde{D}$,
$B_s \rightarrow \phi K_S$, $\tilde{B} \rightarrow \rho K_S$,
$B_S \rightarrow \psi K_S$, $\tilde{B} \rightarrow \phi K_{S}$,
$ \tilde{B} \rightarrow K_S K_S $, $B_s \rightarrow \eta^{'}
\eta^{'}$, $B_s \rightarrow \phi K_S$ and $B_s \rightarrow \psi \phi$

\subparagraph{Tree-dominated decays: $\tilde{B} \rightarrow \pi \pi$.}

The decay $\tilde{B}  \rightarrow \pi \pi $ proceeds through the
quark decay $b \rightarrow u \bar{u} d$, for which both the tree
and the penguin diagrams have CKM parameters of order $\lambda^{3}$.
So  the tree diagram is dominant and
\begin{equation}
\lambda_{\pi\pi} = \frac{q}{p} \cdot \frac{\tilde{A}}{A} \simeq
\frac{V^{*}_{tb}V_{td}}{V_{tb}V^{*}_{td}} \cdot \frac{V_{ub}V^{*}_{ud}}
{V^{*}_{ub}V_{ud}} =
e^{-2i\beta}e^{-2i\gamma} = e^{2i\alpha} \,,
\end{equation}
\begin{equation}
Im(\lambda_{\pi\pi}) \simeq \sin(2\alpha) \,.
\end{equation}
Hadronic uncertainties arise from the small admixture of penguin
contributions and they are expected to be of order 10 percent.

\subparagraph{Tree-forbidden decays: $\tilde{B} \rightarrow \phi
K_{S}$.}

The decay $\tilde{B} \rightarrow \phi K_{S}$ proceeds through
the quark transition $b \rightarrow s\bar{s} s$, i.e. it is
forbidden at tree level and only penguin diagram contributes.  A new
ingredient here is the presence of $K - \tilde{K}$ mixing which adds
a factor
\begin{equation}
(\frac{q}{p})_K \simeq \frac{V_{cs}V^{*}_{cd}}{V^{*}_{cs}V_{cd}} \,.
\end{equation}
in the definition of $\lambda$.
One can find that
\begin{equation}
\lambda_{\phi K_S} = (\frac{q}{p})_B \cdot (\frac{q}{p})_K \cdot
\frac{\tilde{A}}{A} \simeq
\frac{V_{tb}^{*}V_{td}}{V_{tb}V^{*}_{td}}
\cdot \frac{V_{cs}V_{cd}^{*}}{V^{*}_{cs}V_{cd}}\cdot
\frac{V_{tb}V^{*}_{ts}}{V^{*}_{tb}V_{ts}} = e^{-2i\beta} \,,
\end{equation}
\begin{equation}
Im(\lambda_{\phi K_S}) \simeq -\sin(2\beta) \,.
\end{equation}

\subparagraph{Decays with a single weak phase:
$\tilde{B} \rightarrow \psi K_S$.}

The decay $\tilde{B} \rightarrow \psi K_S$ is based on the quark
transition $b \rightarrow c\bar{c} s$, for which the tree diagram is
dominant. One can find that
\begin{equation}
\lambda_{\psi K_S} = -(\frac{q}{p})_B \cdot (\frac{q}{p})_K \cdot
\frac{\tilde{A}}{A} \simeq  - e^{-2i\beta} \,,
\end{equation}
\begin{equation}
Im(\lambda_{\psi K_S}) \simeq \sin(2\beta) \,.
\end{equation}
The hadronic uncertainties for this decay are of the order $10^{-3}$.

Note that the angle $\beta^{'}$ appearing in the CP asymmetries for
$B_s$-meson decay is the analogue of the angle $\beta$ in the
unitarity triangle defined by the relation
\begin{equation}
V_{us}V^{*}_{ub} + V_{cs}V^{*}_{cb} + V_{ts}V^{*}_{tb} = 0 \,.
\end{equation}
Experimentally $|\sin(2\beta^{'})| \le 0.06$.

\paragraph{3.2.5 B-physics at CMS }
\subparagraph{ }

Three experiments are foreseen for the B-physics investigation at
LHC. ATLAS and CMS are two general purpose experiments designed
to look for Higgs boson  and supersymmetric particles. LHC-B \cite{26}
is dedicated experiment for the CP violation study.
Since ATLAS and CMS are designed to look for
particles produced in a very hard collision
detectors cover the central region. For the initial phase of the
LHC operation where the luminosity is around $10^{33}cm^{-2}s^{-1}$,
they intend to do physics with B-mesons. The b-quark events are
triggered by the high transverse momentum $(p_t)$ lepton trigger by
reducing the threshold value.

LHC-B chose the forward geometry due to the following reasons
\cite{27}:

1. The b-quark production is peaked in the forward direction and
in the forward region both b and $\bar{b}$ go to the same direction.
Therefore, a single arm spectrometer with a modest angular coverage
of up to $ \sim 400$ mrad can detect 10 to 20 percent of $b\bar{b}$
events where decay products of the both b-hadrons are in detector
acceptance. This reduces the cost of the detector.

2. B-hadrons produced in the forward direction are faster than those
in the central region. Their average momentum is about 80 Gev,
corresponding to a mean decay length  of $\sim 7 $ mm. Therefore, a
good decay time resolution can be obtained for reconstructed
B-mesons.

3. In the forward region, momenta are mainly carried by the
longitudinal components. Therefore, the threshold value for $p_t$
trigger can be set low for electrons, muons and hadrons around 1.5
Gev. This makes $p_t$ trigger more efficient than in the central
region.

4. Detector can be built in an open geometry which allows easy
installation and maintenance.

5. LHC-B is only the detector capable of separating kaons from
pions in all necessary phase space.

Here we briefly describe the B-physics potential of CMS.
ATLAS B-physics potential is similar to CMS one.
As it has been mentioned before the
main CP violation prediction for B systems is the inequality
$\Gamma(B^{0} \rightarrow f) \neq \Gamma(\bar{B}^{0} \rightarrow
\bar{f})$.  The decay rate asymmetry
\begin{equation}
A =
\frac{\Gamma(B^0 \rightarrow f) - \Gamma(\bar{B}^0 \rightarrow
\bar{f})} {\Gamma(B^0 \rightarrow f) + \Gamma(\bar{B}^0 \rightarrow
\bar{f})} \sim \sin(2\Phi) \,.
\end{equation}
depends only on CP-violating angles of triangle
$\alpha, \gamma, \beta = \Phi$.  At LHC the cross section
$pp \rightarrow b\bar{b} +...$ is expected to be in the range
$0.08 - 0.6$ mb. The most promising channels for the
search for the CP-violation are:

$B^0_d \rightarrow (\psi \rightarrow \mu^+ \mu^-) + (K^0_s
\rightarrow \pi^{+} \pi^{-})$ and $B^0_d \rightarrow \pi^+ \pi^-$.

The decay $B^0_d \rightarrow \psi K^0_S$ with $\psi \rightarrow
\mu^{+} \mu^{-}$ and $K^0_S \rightarrow \pi^{+}\pi^{-}$ is the
most appropriate channel to measure the angle $\beta$. To tag
the $B^0_d$, the associated b-hadron is required to decay into
muon + X. The time integrated asymmetry A is:
\begin{equation}
A = \frac{N^+ - N^-}{N^+ +N^-} = D\cdot \frac{x_d}{1+x^2_d}\cdot
\sin(2\beta)\,,
\end{equation}
where $N^+$ and $N^-$ are the number of events with positively and
negatively charged tagging muons, D is the dilution factor and
$\frac{x_d}{1+x^2_d}$ is the time-integration factor. The expected
number of events for an integrated luminosity L is:
\begin{eqnarray}
&&N  = 2 \times L \times \sigma_{b\bar{b}} \times P(\bar{b} \rightarrow
B^0_d) \times Br(B^0_d \rightarrow \psi K^0_S) \times
Br(\psi \rightarrow \mu^{+}\mu^{-}) \\ \nonumber
&&\times Br(K^0_S \rightarrow
\pi^{+}\pi^{-}) \times Br(b \rightarrow \mu) \times A_{trig} \times
\epsilon  \,,
\end{eqnarray}
where $A_{trig}$ is the trigger acceptance and $\epsilon$ is the
efficiency of the selection cuts. Typical values of the branching
ratios are: $P(\bar{b} \rightarrow B^0_d) =0.4$;
$Br(B^0_d  \rightarrow \psi K^0_S) =3.3 \times 10^{-4}$; $Br(\psi
\rightarrow \mu^{+}\mu^{-}) = 0.0597$; $Br(K^0_S \rightarrow \pi^{+}
\pi^{-}) = 0.6861$; $Br(b \rightarrow \mu) =0.105$. Before
trigger-acceptance and data-selection cuts, the number of produced
signal events is $5.6 \times 10^{6}$ for $10^{4}$ $pb^{-1}$. The
measured asymmetry A is affected by dilution factors $D \simeq 0.47$.
The signal to background ratio is estimated to be $S/b \approx
10:1$ and the mixing angle $\sin(2\beta)$ accuracy determination is
$\approx 0.05$ for $L_t = 10^{4}$ $pb^{-1}$.

The $B^0_d \rightarrow \pi^{+} \pi^{-}$ decay is a good channel to
determine the angle $\alpha$ of the unitarity triangle. For $L_t =
10^{4}$ $pb^{-1}$ the sensitivity to the triangle angle $\alpha$
determination is estimated to be:

$\delta(\sin(2\alpha))$ =  $0.057 \pm 0.018$ \,.

CP-violation in $B^0_d$ and $\tilde{B}^0_d$ decays gives rise to
different decay rates  with a time dependence. The asymmetry can be
measured as a function of proper time $t/\tau$ in units of the
$B^0_d$ lifetime $\tau$:
\begin{equation}
A(t/\tau) = \frac{N^- - N^+}{N^- + N^+} = D\cdot \eta_{CP} \cdot
\sin(2\varphi_{i})\cdot \sin(x_dt/\tau) \,,
\end{equation}
where  $N^+$ and $N^-$ are the numbers of events with positively
and negatively charged tagging muons, D is the dilution factor,
$\eta_{CP}$ is the CP-parity of the final state f and
$\varphi_{i}$ is the angle of the unitarity triangle.
It is possible
to measure the secondary vertex in the transverse plane to determine
time-dependent asymmetry and hence to determine the CP-violating
angles $\alpha$ and $\beta$ \cite{1}. The accuracy in the
determination of the angles $\alpha$ and $\beta$ is similar to the
accuracy for the case the time integrated asymmetry considered
previously.

Very important  task of B physics at LHC is the
observation of $B_s^0 - \bar{B}^0_s$ oscillations but it could be
very difficult if $x_s$ is large. The current limits on the value
of $x_s$ are $5.6 \leq x_s \leq 33.2$ \cite{28}. The possibility
to discover $B_s^0 - \bar{B}^0_s$ oscillations at LHC have been
studied in refs. \cite{29}.

Other interesting task is the search for $B^0_s
\rightarrow \mu^+\mu^-$ rare decay. This decay is forbidden at tree
level. In standard model the branching ratio of this decay is
expected  to be $\approx 2\times 10^{-9}$ \cite{30}. It appears that
at CMS the upper limit on this branching ratio can be set
$1.4\times10^{-9}$ at a 90 percent C.L. for $L_t = 3\times 10^{4}$
$pb^{-1}$ \cite{1}.

\paragraph{3.2.6 Heavy ion physics}

\subparagraph{ }

Quantum chromodynamics has already proved itself to be a reliable
theory when dealing with quark and gluon interactions at short
distances. The QCD vacuum shows properties which are similar
to those of a superconductor. It has a critical temperature
beyond which it becomes transparent to colour. This temperature is
found to be of the order 150 - 200 Mev \cite{31}. Consider a system
consisting of a set of hadrons in the vacuum and compress it. The
hadrons are initially too far to overlap. As the density increases
they start overlapping among themselves. We thus go from a dilute
to a very dense hadron gas. With increasing compression the hadrons
merge among one another. The small vacuum bubbles separately
associated with each individual hadron should eventually fuse into
one big bubble within which quarks would freely move over distances
much larger than those offered by the hadrons in which they were
first confined. Since in high energy collisions new particles
are created out of the collision energy, we reach the same
conclusions if the initial dilute hadron gas is heated instead
of being merely compressed. In heavy ion collisions we have a
complicated mixture of compression and heating. Computer calculations
support the naive picture according to which at low temperature we
have hadron gas and at
high temperature we have quark-gluon plasma \cite{31}.
Naively we can expect the following picture in heavy ion
collisions at high energies \cite{31}:

1. An initial phase during which many collisions occur at the parton
level. A large number of energetic partons are formed. The system is
still far from any equilibrium. A large amount of entropy is
released.

2. A thermalized phase, obtained through the scattering of the many
partons. The temperature is very high. The system is a quark-gluon
plasma, at least in some localized regions.

3. A mixed phase, obtained as the plasma cools with parts of it
hadronizing as the temperature goes through its critical value.

4. A thermalized, dense, hadron gas from which hadrons eventually
escape as their mean free path exceeds the size of the system. This
is what is referred to as freeze out.

However all the processes occur in a very short time. When we look
at the dominantly produced hadrons we merely integrate over the whole
evolution. So any evidence for a quark-gluon plasma is averaged out
the past history. In other words after the transition from
quark-gluon plasma to hadron phase the system can "forget" about
previous phase.  The most striking signal associated
with quark-gluon plasma is the suppression of the charmonium
and upsilonium formation \cite{32}.

The physical picture leading to the $\psi$ suppression due to
quark deconfinement in nuclear collisions is quite simple.
Within a deconfining medium like QGP(quark-gluon plasma),
quarks cannot bind to form hadrons. Heavy charm quark-antiquark
pairs which form $\psi$ are produced by hard, prethermal
interactions at a very early stage of the collision. In a
deconfining medium $c$ and $\bar{c}$ just fly apart. Moreover,
within the equilibrated QGP the production of additional
thermal $c$ quarks is strongly suppressed by a factor
$\exp(-m_c/T_c) \simeq 0.6 \times 10^{-3}$ with a mass of charm quark
$m_c = 1.5$ Gev and $T_c = 0.2$ Gev. Consequently $\psi$ production
will be suppressed in the presence of QGP. In atomic physics, the
screened binding potential between two charges is given by $V = V_o
\exp(-r/r_D)$, where $r_D$ is the Debye radius. When $r_D \approx
r_B$ the valence electrons are liberated owing to charge screening
and the insulator changes into a conductor. Similarly, in QCD
the potential between two coloured quarks is given by $V = V_0
\exp(-r/r_D)$ because of colour screening, and if the density of colour
charges is sufficient to make $r_D \leq r_H$ (hadron radius), colour
insulators of hadrons change into colour-conducting phase, i.e. QGP.
It means that the long range confining phase of the potential gets
screened, i.e. we have quark deconfinement. We have to know the
$\psi$ radius and the screening radius $r_D$ at a given temperature.
Only for $r_D \leq r_{\psi}$ we can expect $\psi$ suppression.
It has been found \cite{31} that $r_D = r_H$ occurs near the
transition temperature $T_c$ for $\psi^{'}$  and
$T/T_c \geq 1.3$ for $\psi $. Recent data of NA50 \cite{33} support
this picture. It should be noted however that $\psi$ suppression
can also be explained either on the basis of absorption processes in
a dense hadron gas or on the basis of conventional nuclear effects
\cite{34}. For LHC energies with the center of mass
energy around 6 Tev per nucleon and with the energy density
$\epsilon \approx 9$ $Gev/fm^3$ all heavy quark bound states
except for $\Upsilon(1S)$ will be suppressed by color
screening.

The main goal of LHC heavy ion program is the search for quark-gluon
plasma. There will be special detector ALICE \cite{35} dedicated for the
heavy ion investigation at LHC. Here we briefly describe the
"heavy ion potential" of CMS \cite{1}.
The most promising signature here is the measurement for the
production rates of the bottonium states for different nuclei and
different kinematics. The most interesting signature is the
measurement of the cross section of $\Upsilon$ state as a
function of $p^{\Upsilon}_{T}$ since no suppression due to Debye
colour screening is expected \cite{31,32}. As it has been mentioned before
the typical prediction of the existence of quark-gluon plasma
in this case is the suppression of $\Upsilon(2S)$
and $\Upsilon(3S)$ relative to $\Upsilon(1S)$
yield. Simulation studies were performed  for
$^{16}O$, $^{40}Ca$, $^{97}Nb$ and $^{208}Pb$ ion collisions.
The $\psi$ and $\Upsilon$ resonances were generated using $p_t$
and $y$ distributions extrapolated from experimental results
obtained at lower $\sqrt{s}$. The total acceptance for
$\Upsilon \rightarrow \mu \mu $ has been found $\approx 0.33$,
and for $\psi$ it is around 0.06. If we restrict ourselves to the
barrel region, where muons with $p_t \geq 4$ Gev can be detected,
the acceptance for $\Upsilon$ decreases to 0.05 and for $\psi$ it
vanishes. It has been found that the mass resolution for
$\Upsilon$ is around 40 Mev. The cross sections for $\Upsilon$
production in $A-A$ collisions were obtained multiplying the
cross sections for $pp$ reaction at the same c.m.s. energy by
factor $A^{2\alpha}$  with $\alpha = 0.95$. The main dimuon
background is due to uncorrelated muon pairs from
$\pi$ and $K$ decays. The signal-to-background ratios have been
calculated in the mass band $M(\Upsilon) \pm 50$ Mev, for central
collisions of  Pb, Ca, Nb and O beams. The results depend rather
strongly on the value of $\sigma(pp \rightarrow \Upsilon)$ and
on the charged  particle multiplicity $dn^{\pm}/dy$
per unit rapidity in the central collision. The main conclusion
from these studies is that CMS will be able to detect reliably
$\Upsilon$, $\Upsilon^{'}$ and  $\Upsilon^{''}$
resonances.

Other interesting prediction testable at CMS is the jet quenching in
quark gluon plasma. As jets are produced early in a
collision, they propagate in the plasma, and interact strongly
before they escape. Hence, they carry information about deconfined
hadronic matter. The problems of jet finding at CMS with the large
transverse energy $(E_t)$ flow in central collisions were studied, to
determine the possibilities of observing jet quenching using CMS
\cite{36}. The high $E_t$ jet production cross-section in $Pb-Pb$
collisions was estimated in the rapidity region $|y| \leq 1.5$, using
the jet cross-section for $pp$ interactions, as evaluated by PYTHIA,
and an $A^{2\alpha}$ dependence for the $A-A$ colllisions. The effect
of energy losses in dense nuclear matter changing the high $E_t$
jet production cross section was not taken into account. A central
event with two high $E_t$ jets from a proton-proton collision was
superimposed on the ion collision.  The main conclusion from these
studies is that jets may be reliably reconstructed in the CMS
detector for $Pb-Pb$ central
collisions with $E_t \ge 100$ Gev \cite{36}.

\subsection{Supersymmetry search within MSSM}

\paragraph{3.3.1 MSSM model}

\subparagraph{ }

Supersymmetry is a new type of symmetry that relates properties of
bosons to those to fermions \cite{37}. It is the largest symmetry of
the S-matrix. Locally supersymmetric theories necessarily incorporate
gravity \cite{38}. SUSY is also an essential ingredient of superstring
theories \cite{39}. The recent interest in supersymmetry is due to
the observation that measurements of the gauge coupling constants at
LEP1 are in favour of the Grand Unification in a supersymmetric
theory with superpartners of ordinary particles
which are lighter than $O(1)$ Tev. Besides supersymmetric electroweak
models offer the simplest solution of the gauge hierarchy problem
\cite{37}. In real life supersymmetry has to be broken and
to solve the gauge hierarchy problem the masses of superparticles
have to be lighter than $O(1)$ Tev. Supergravity
gives natural explanation of the supersymmetry breaking \cite{38},
namely, an account of the supergravity breaking in hidden sector
leads to soft supersymmetry breaking in observable sector.
The simplest supersymmetric generalization of the standard model
is the Minimal Supersymmetric Standard Model (MSSM). It is
supersymmetric theory based on standard $SU_{c}(3) \otimes SU_L(2)
\otimes U(1)$ gauge group with electroweak symmetry spontaneously
broken via vacuum expectation values of two different Higgs
doublets. The MSSM consists of taking the standard model and adding
the corresponding supersymmetric partners. It should be stressed that
in addition the MSSM contains two hypercharges $Y = \pm 1$ Higgs
doublets, which is the minimal structure for the Higgs sector of
an anomaly-free supersymmetric extension of the Standard Model.
The supersymmetric electroweak models also require at least two
Higgs doublets to generate masses for both "up"-type and
"down"-type fermions.
The renormalizable superpotential determines the Yukawa
interactions of quarks and leptons and preserves global
$B - L$. Here B is the baryon number and L is the lepton
number. Note that the most general expression for the
effective superpotential contains renormalizable terms
violating $B - L$ that can lead to the problems with
proton decay. To get rid of such dangerous terms in the
superpotential R-parity conservation of the theory is
postulated. Here $R = (-1)^{3(B - L) +2S}$ for a particle
of spin S. This formula means that all ordinary standard model
particles have $R = 1$, whereas the corresponding supersymmetric
partners have $R = - 1$. The R-parity conservation has a crusial
impact on supersymmetric phenomenology. An important consequence
of R-parity conservation is that the lightest supersymmetric
particle (LSP) is stable.  The cosmological constraints imply
that  LSP is weakly-interacting electrically neutral and
coloruless particle. Other important consequences of R-parity
conservation is that at supercolliders  superparticles
have to be produced in pairs, therefore at least two LSP have
eventually be produced at the end of the decays of heavy
unstable supersymmetric particles. Being weakly interacting
particle LSP escapes detector registration, therefore
the classic signature for R-parity conserving supersymmetric
theories is the transverse missing energy due to the
LSP escape. Note that at present there are no deep theoretical
arguments in favour of R-parity conservation. There are
models with R-parity violation \cite{40}. Models with
R-parity violation break $B-L$ number and are strongly constrained
by existing experimental data. In such models LSP is unstable,
supersymmetric particles can be singly produced and in general
the signature related with the transverse missing energy is lost.

The superpotential of MSSM has the form
\begin{equation}
W = h^u_{ij}U^c_iQ_j H_2 + h^d_{ij}D^c_iQ_j H_1
  + h^l_{ij}E^c_i L_jH_1 -
\mu H_1 H_2 \,,
\end{equation}
where $i,j$ are summed over 1,2,3 and $Q_j$, $U^c_a$ ,
$D^c_b$ denote $SU(2)$ doublet and $SU(2)$  singlet quark superfields,
$L^i$, $E^c_i$ are the $SU(2)$ doublet and $SU(2)$ singlet lepton
superfields and $H_1$, $H_2$ denote the two Higgs superdoublets,
$h^u_{ij}$, $h^d_{ij}$, $h^l_{ij}$ are the Yukawa coupling constants.  In
MSSM supersymmetry is softly broken at some high scale $M$ by generic
soft terms
\begin{eqnarray}
&&-L_{soft} =
m_{0}(A^u_{ij}U^c_iQ_jH_2 +A^d_{ij}D^c_i G_j H_1 + \\ \nonumber
&&A^l_{ij}E^c_iL_jH_1 + h.c.) +
(m^2_q)_{ij}Q^+_iQ_j  +(m^2_u)_{ij}(U^c_i)^+U^c_j \\ \nonumber
&&+ (m^2_d)_{ij}(D^c_i)^+ D^c_j + (m^2_l)_{ij}(L^c_i)^+ L^c_j +
(m^2_e)_{ij}(E^c_i)^+ \\ \nonumber
&&E^c_j + m^2_1 H_1H^+_1
+ m^2_2H_2H^+_2
+ (B{m_{0}}^2 H_1H_2 + h.c.) +  \frac{1}{2}m_a (\lambda_{a}
\lambda_{a})  \,.
\end{eqnarray}

In most analysis the mass terms are supposed to be diagonal at
$M_{GUT} \approx 2\cdot 10^{16}$ Gev scale and gaugino and
trilinear mass terms are supposed to be universal at $M_{GUT}$
scale, namely at GUT scale:
\begin{equation}
A^u_{ij}(M_{GUT}) = Ah^u_{ij}(M_{GUT}), \;
A^d_{ij}(M_{GUT}) = Ah^d_{ij}(M_{GUT}), \; A^l_{ij}(M_{GUT}) = A
h^l_{ij}(M_{GUT}),
\end{equation}
\begin{eqnarray}
&&(m^2_q)_{ij}(M_{GUT}) = (m^2_u)_{ij}(M_{GUT}) = (m^2_d)_{ij}
(M_{GUT}) =  \\ \nonumber
&&(m^2_l)_{ij}(M_{GUT}) = (m^2_e)_{ij}(M_{GUT}) =
\delta_{ij} m^2_1(M_{GUT}) = \delta_{ij} m^2_2(M_{GUT}) =
\delta_{ij} m_0^2 \,,
\end{eqnarray}
\begin{equation}
m_1(M_{GUT}) = m_2(M_{GUT}) = m_3(M_{GUT}) = m_{1/2} \,.
\end{equation}
Note that it is more appropriate to impose boundary conditions
not at GUT scale but at Planck scale $M_{PL} = 2.4\cdot 10^{18}$ Gev.
An account of the renormalization effects between PLANCK scale and
GUT scale can drastically change the features of the spectrum.
For instance, if we assume that the physics between Planck scale and
GUT scale is described by SUSY SU(5) model then an account of
the evolution between PLANCK and GUT scales \cite{41,42} changes
qualitatively the spectrum of sleptons for $m_{0} \ll m_{1/2}$ \cite{42}.
So in MSSM we have unknown  soft supersymmetry breaking
parameters $m_0$, $m_{1/2}$, $A$, $B$ plus we
have unknown parameter $\mu$ in the superpotential.
The renormalization group equations for soft SUSY
breaking parameters in neglection of all Yukawa coupling constants
except top-quark Yukawa in one loop approximation read \cite{43}
\begin{equation}
\frac{d\tilde{m}^2_L}{dt} = (3\tilde{\alpha}_2M^2_2 +\frac{3}{5}
\tilde{\alpha}_1 M^2_1) \,,
\end{equation}
\begin{equation}
\frac{d\tilde{m}^2_E}{dt} = (\frac{12}{5}\tilde{\alpha}_1M^2_1) \,,
\end{equation}
\begin{equation}
\frac{d\tilde{m}^2_Q}{dt} = (\frac{16}{3}\tilde{\alpha}_3M^2_3 +
3\tilde{\alpha}_2M^2_2 + \frac{1}{15} \tilde{\alpha}_1 M^2_1) -
\delta_{i3}Y_t(\tilde{m}^2_Q +\tilde{m}^2_U + m^2_2 + A^2_tm^2_0 -
\mu^{2}) \,,
\end{equation}
\begin{equation}
\frac{d\tilde{m}^2_U}{dt} = (\frac{16}{3}\tilde{\alpha}_3M^2_3 +
\frac{16}{15}\tilde{\alpha}_1M^2_1) - \delta_{i3}2Y_t(\tilde{m}^2_Q
+\tilde{m}^2_U + m^2_2 +A^2_tm^2_0 - \mu^{2}) \,,
\end{equation}
\begin{equation}
\frac{d\tilde{m}^2_D}{dt} = (\frac{16}{3}\tilde{\alpha}_3M^2_3 +
\frac{4}{15}\tilde{\alpha}_{1}M^2_1) \,,
\end{equation}
\begin{equation}
\frac{d\mu^{2}}{dt} = 3(\tilde{\alpha}_2 +\frac{1}{5}\tilde{\alpha}_1
-Y_t)\mu^{2} \,,
\end{equation}
\begin{equation}
\frac{dm^2_1}{dt} = 3(\tilde{\alpha}_2M^2_2 +
\frac{1}{5}\tilde{\alpha}_1M^2_1)
  + 3(\tilde{\alpha}_2 + \frac{1}{5}
\tilde{\alpha}_1 - Y_t)\mu^{2} \,,
\end{equation}
\begin{equation}
\frac{dm^2_2}{dt} =  3(\tilde{\alpha}_2M^2_2
+\frac{1}{5}\tilde{\alpha}_1M^2_1) + 3(\tilde{\alpha}_2 + \frac{1}{5}
\tilde{\alpha}_1)\mu^{2}
-3Y_t(\tilde{m}^2_Q + \tilde{m}^2_U + m^2_2 + A^2_t m^2_0) \,,
\end{equation}
\begin{equation}
\frac{dA_t}{dt} = -(\frac{16}{3}\tilde{\alpha}_3\frac{M_3}{m_0} +
3\tilde{\alpha}_2\frac{M_2}{m_0} + \frac{13}{15} \tilde{\alpha}_1\frac
{M_1}{m_0})) - 6Y_tA_t \,,
\end{equation}
\begin{equation}
\frac{dB}{dt} = -3(\tilde{\alpha}_2\frac{M_2}{m_0} +
\frac{1}{5} \tilde{\alpha}_1 \frac{M_1}{m_0}) - 3Y_tA_t \,,
\end{equation}
\begin{equation}
\frac{dM_i}{dt} = -b_i\tilde{\alpha}_iM_i \,,
\end{equation}
\begin{equation}
b_1 =\frac{33}{5}, \; b_2 =1,\;b_3=-3 \,.
\end{equation}
Here $\tilde{m}_U$, $\tilde{m}_D$, $\tilde{m}_E$ refer to the masses
of the superpartner of the quark and lepton singlets, while
$\tilde{m}_Q$ and $\tilde{m}_L$ refer to the masses of the isodoublet
partners; $m_1$, $m_2$, $m_3$ and $\mu$ are the mass partners of
the Higgs potential, $A$ and $B$ are the couplings of the $L_{soft}$
as defined before; $M_i$ are the gaugino masses before mixing. The
renormalization group equation for the top
Yukawa coupling constant has the form
\begin{equation}
\frac{dY_t}{dt} =  Y_t(\frac{16}{3}\tilde{\alpha}_3 +3
\tilde{\alpha}_{2} +\frac{13}{15} \tilde{\alpha}_{1}) - 6Y^2_t \,,
\end{equation}
while the RG equations for the gauge couplings are
\begin{equation}
\frac{d\tilde{\alpha}_i}{dt} = -b_i\tilde{\alpha}^2_i \,.
\end{equation}
Here
\begin{equation}
\tilde{\alpha}_i = \frac{\alpha_{i}}{4\pi},\;
Y_t =\frac{h^2_t}{16\pi^{2}},\; t = \ln{(\frac{M^2_{GUT}}{Q^2})} \,,
\end{equation}
and the top Yukawa coupling $h_t$ is related to the running top
mass by the relation
\begin{equation}
m_t = h_t(m_t)\frac{v}{\sqrt{2}}\sin{\beta} \,.
\end{equation}
The boundary conditions at $Q^2 = M^2_{GUT}$ are
\begin{equation}
\tilde{m}^2_Q = \tilde{m}^2_U = \tilde{m}^2_D = \tilde{m}^2_E =
\tilde{m}^2_L = m^2_0 \,,
\end{equation}
\begin{equation}
\mu = \mu_{0}; \; m^2_1 = m^2_2 = \mu^{2}_{0} +m^2_0; \;
m^2_3 = B\mu_{0} m_0 \,,
\end{equation}
\begin{equation}
M_i = m_{1/2}, \; \tilde{\alpha}_{i}(0) = \tilde{\alpha}_{GUT}; \;
i =1,2,3 \,.
\end{equation}
For the gauginos of the $ SU(2) \otimes U(1)$ gauge group one has to
consider the mixings with the Higgsinos. The mass terms in the full
lagrangian are \cite{44}
\begin{equation}
L_{Gaugino-Higgsino} = -\frac{1}{2}
   M_3 \bar{\lambda}^a_3 \lambda^{a}_{3}
   -\frac{1}{2} \bar{\chi} M^{(0)} \chi -
   (\bar{\psi}M^{(c)}\psi +h.c.) \,,
\end{equation}
where $\lambda^{a}_{3}$ are the 8 Majorana gluino fields, and
\begin{equation}
\chi = \left( \begin{array}{cccc}
\tilde{B}^0\\
\tilde{W}^3\\
\tilde{H}^0_1\\
\tilde{H}^0_2
\end{array} \right) \,,
\end{equation}
\begin{equation}
\psi = \left( \begin{array}{cc}
\tilde{W}^+\\
\tilde{H}^+
\end{array} \right) \,,
\end{equation}
are the Majorano neutralino and Dirac chargino fields. The mass
matrices are:
\begin{equation}
M^{(0)} = \left( \begin{array}{cccc}
M_1 & 0 & -A & B\\
0 & M_2 & C & -D \\
-A & C & 0 & -\mu\\
B & -D & -\mu & 0
\end{array} \right) \,,
\end{equation}
\begin{equation}
M^{(c)} = \left( \begin{array}{cc}
M_2 & \sqrt{2}M_W \sin{\beta} \\
\sqrt{2}M_W\cos{\beta} & \mu
\end{array} \right) \,,
\end{equation}
where:
\begin{equation}
A = M_Z\cos{\beta}\sin{\theta_{W}}, \; B = M_Z
\sin{\beta}\sin{\theta_{W}} \,,
\end{equation}
\begin{equation}
C =M_Z\cos{\beta}\cos{\theta_{W}}, \; D =
M_Z\sin{\beta}\cos{\theta_{W}} \,.
\end{equation}
After the solution of the corresponding renormalization group
equations for $\alpha_{GUT} = \frac{1}{24.3}$, $M_{GUT} =
2.0\cdot10^{16}$ Gev, $\sin^{2}{\theta_{W}} =0.2324$ and
$A_t(0) = 0$ one finds the numerical formulae for squark and
slepton square masses \cite{45}
\begin{equation}
\tilde{m}^2_{E_{L}}(t =66) = m^2_0 + 0.52m^2_{1/2} -0.27\cos{2\beta}
M^2_Z \,,
\end{equation}
\begin{equation}
\tilde{m}^2_{\nu_{L}}(t =66) =m^2_0 + 0.52m^2_{1/2}
+0.5\cos{2\beta}M^2_Z \,,
\end{equation}
\begin{equation}
\tilde{m}^2_{E_{R}}(t=66) = m^2_0 +0.15m^2_{1/2}
-0.23\cos{2\beta}M^2_Z  \,,
\end{equation}
\begin{equation}
\tilde{m}^2_{U_{L}}(t=66) = m^2_0 + 6.5m^2_{1/2}
+0.35\cos{2\beta}M^2_Z  \,,
\end{equation}
\begin{equation}
\tilde{m}^2_{D_{L}}(t=66) = m^2_0 + 6.5m^2_{1/2} -
0.42\cos{2\beta}M^2_Z \,,
\end{equation}
\begin{equation}
\tilde{m}^2_{U_{R}}(t=66) = m^2_0 +6.1m^2_{1/2}
+0.15\cos{2\beta}M^2_Z  \,,
\end{equation}
\begin{equation}
\tilde{m}^2_{D_{R}}(t=66) = m^2_0 +6.0m^2_{1/2}
-0.07\cos{2\beta}M^2_Z  \,,
\end{equation}
\begin{equation}
\tilde{m}^2_{b_{R}}(t=66) = \tilde{m}^2_{D_{R}}  \,,
\end{equation}
\begin{equation}
\tilde{m}^2_{b_{L}}(t=66) = \tilde{m}^2_{D_{L}} - 0.49m^2_0
-1.21m^2_{1/2}  \,,
\end{equation}
\begin{equation}
\tilde{m}^2_{t_{R}}(t=66) = \tilde{m}^2_{t_{R}}(t=66) =
\tilde{m}^2_{U_{R}}(t=66) +m^2_t -0.99m^2_0 -2.42m^2_{1/2}  \,,
\end{equation}
\begin{equation}
\tilde{m}^2_{t_{L}}(t=66) = \tilde{m}^2_{U_{L}}(t=66) + m^2_t - 0.49m^2_0
-1.21 m^2_{1/2} \,.
\end{equation}
After mixing the mass eigenstates of the stop matrix are:
\begin{eqnarray}
&&\tilde{m}^2_{t_{1,2}}(t=66) \approx
\frac{1}{2}[0.5m^2_0 +9.1m^2_{1/2} +2m^2_t +
0.5\cos{2\beta}M^2_Z] \\ \nonumber
&&\mp \frac{1}{2}[(1.5m^2_{1/2} + 0.5 m^2_0 + 0.2\cos{(2\beta)}M^2_Z)^2
+4m^2_t(A_tm_o -\mu/\tan{\beta})^2]^{1/2} \,.
\end{eqnarray}
The gauginos and Higgsinos have similar quantum numbers which causes
a mixing between the weak interaction eigenstates and the mass
eigenstates. The two chargino eigenstates $\chi^{\pm}_{1,2}$ are:
\begin{equation}
M^2_{1,2} = \frac{1}{2}[M^2_2 + \mu^{2} + 2M^2_W] \mp
\frac{1}{2}[(M^2_2 -\mu^{2})^2 +
4M^4_W\cos^2{2\beta} + 4M^2_W(M^2_2 +\mu^{2} +
2M_2\mu\sin{2\beta})]^{1/2} \,,
\end{equation}
where at GUT scale the masses of the gaugino fields of the $SU(3)$,
$SU_L(2)$ and $U(1)$ groups are equal to $m_{1/2}$. The eigenvalues of
the $4\times 4$ neutralino mass matrix can be solved by a numerical
diagonalization. If the parameter $\mu$ is much larger than $M_1$
and $M_2$, the mass eigenstates become
\begin{equation}
\chi^{0}_{i} = [\tilde{B}, \tilde{W}_{3},
\frac{1}{\sqrt{2}}(\tilde{H}_{1} - \tilde{H}_{2}),
\frac{1}{\sqrt{2}}(\tilde{H}_{1} + \tilde{H}_{2})] \,
\end{equation}
with eigenvalues $|M_1|$, $|M_2|$, $|\mu| $ and $|\mu|$ respectively
( the bino and neutral wino do not mix with each other nor with the
Higgsino eigenstates).

The tree level Higgs potential in MSSM has the form
\begin{equation}
V_{0}(H_1,H_2) = m^2_1|H_1|^2 + m^2_2|H_2|^2 - m^2_3(H_1H_2 + h.c)
+\frac{g^2_2 + \tilde{g}^2_1}{8}(|H_1|^2 - |H_2|^2)^2 + \frac{g^2_2}{2}
|H^{+}_1H_2|^2 \,,
\end{equation}
where $\tilde{g}^2_1 = \frac{3}{5}g^2_1$.

The minimization of the effective potential $V_0(H_1,H_2)$ leads to
the equations:
\begin{equation}
v^2 = \frac{8(m^2_1-m^2_2\tan^2{\beta})}{(g^2_2 +\tilde{g}^2_1)
(\tan^2{\beta} -1)} \,,
\end{equation}
\begin{equation}
\sin{2\beta} = \frac{2m^2_3}{m^2_1 + m^2_2} \,.
\end{equation}
 After the diagonalization of the corresponding mass matrices
CP-odd neutral Higgs boson $A(x)$ acquires a mass $m^2_A = m^2_1 +
m^2_2$, charged Higgs boson $H^{+}(x)$ acquires a mass $m^2_{H^{+}} =
m^2_{A} + M^2_W$ and CP-even Higgs bosons $H(x)$ and $h(x)$ have
masses
\begin{equation}
m^2_{H,h} = \frac{1}{2}[m^2_A +M^2_Z \pm
\sqrt{(m^2_A + M^2_Z)^2 - 4m^2_AM^2_Z\cos^2{2\beta}}] \,,
\end{equation}
where $<H_1> = v_1 = \frac{v\cos{\beta}}{\sqrt{2}}$,
$<H_2> = v_2 = \frac{v\sin{\beta}}{\sqrt{2}}$ ,
$\tan{\beta} = \frac{v_2}{v_1}$.
At tree level we have the following mass relations:
\begin{equation}
m^2_h + m^2_H = m^2_A + M^2_Z \,,
\end{equation}
\begin{equation}
m_h \leq m_A \leq m_H \,,
\end{equation}
\begin{equation}
m_h \leq M_Z|\cos{2\beta}| \leq M_Z \,.
\end{equation}
Therefore at tree level the lightest Higgs boson is lighter than the
Z-boson. However the radiative corrections due to big top quark
Yukawa coupling constant increase the mass of the lightest Higgs
boson in MSSM \cite{46}. The upper limit on the Higgs
boson mass in MSSM depends
on the value of the top quark mass and on the value of stop
quark masses. For instance, for $m_t = 175 $ Gev and stop quark
masses lighter than 1 Tev the Higgs boson mass as it follows
from the table 1. has to be lighter than 117 Gev. From the vacuum
stability bound ($k = 0$ in the table 1.) we find that if
standard model is  valid for energies up to the $10^{6}$ Gev then the mass
of the Higgs boson has to be heavier than 121 Gev for $m_t = 175$ Gev.
Therefore for $m_t = 175$ Gev we find that the Higgs
boson mass predictions for SM
and MSSM lie in different mass intervals (in MSSM Higgs boson is
relatively light, whereas in SM it is relatively heavy). So the exact
knowledge of the Higgs boson mass allows to distinguish between SM and
MSSM \cite{47}. For instance, the Higgs boson discovery at LEP2 will
be very powerful evidence in favour of the low energy broken
supersymmetry. After the solution of the corresponding equations for
the determination of nontrivial electroweak potential the number
of unknown parameters is decreased by 2. At present
more or less standard choise of free parameters in MSSM includes
$m_0$, $m_{1/2}$, $\tan{\beta}$, $A$ and sign($\mu$).

\subparagraph{Superparticle cross sections.}

At LHC sparticles can be produced via the following tree level
reactions \cite{48}:

 i. $gg,qq, qg \rightarrow \tilde{g} \tilde{g},\, \tilde{g}
\tilde{q}, \, \tilde{q} \tilde{q}$ ,

ii. $qq, gq \rightarrow \tilde{g} \chi^0_{i},\, \tilde{g}
\chi^{\pm}_i,\, \tilde{q}\chi^0_i,\,\tilde{q} \chi^{\pm}_i$ ,

iii. $qq \rightarrow \chi^{\pm}_i \chi^{\mp}_j,\, \chi^{\pm}_i
\chi^0_j,\, \chi^0_i \chi^0_j $ ,

iiii. $qq \rightarrow \tilde{l} \tilde{\nu}, \, \tilde{l} \tilde{l},
\, \tilde{\nu} \tilde{\nu}$ ,

The Higgs bosons of the MSSM can be produced via direct s-channel
subprocess:

iiiii. $qq, gg \rightarrow h,\, H,\, A,\, H^{\pm}H^{\mp}$ .

It is straightforward to calculate the elementary (tree level)
cross sections for the production of superparticles in collisions of
quarks and gluons. Here following E.Eichten et al., \cite{4}
we collect the main formulae for elementary cross sections.

The differential cross section of the production of two gauge
fermions in quark-antiquark collisions is

\begin{eqnarray}
&&\frac{d\sigma}{dt}(q \bar{q}^{'} \rightarrow gaugino1 +
gaugino2) = \\ \nonumber
&&\frac{{\pi}}{s^2}
[ A_s \frac{(t - m^2_2)(t-m^2_1) + (u -m^2_1)(u -m^2_2) +2s m_1m_2}
{(s -M^2_s)^2} +
A_t \frac{(t-m^2_1)(t-m^2_2)}{(t-M^2_t)^2} + \\ \nonumber
&&A_u \frac{(u-m^2_1)(u-m^2_2)}{(u-M^2_u)^2} +
A_{st} \frac{(t-m^2_1)(t-m^2_2) +
m_1m_2s}{(s - M^2_s)(t-M^2_t)} + \\ \nonumber
&&A_{tu} \frac{m_1m_2s}{(t-M^2_t)(u-M^2_u)}
+ A_{su} \frac{(u-m^2_1)(u-m^2_2) +m_1m_2s}{(s-M^2_s)(u-M^2_u)}] \,,
\end{eqnarray}
where $m_1$ and $m_2$ are the masses of the produced gauginos,
$M_s$, $M_t$ and $M_u$ are the masses of the particles
exchanged in the s,t, and u channels respectively.
The coefficients $A_x$ are given in refs. \cite{47}. For instance,
for the case of the gluino pair
production in quark-antiquark collisions the coefficients $A_x$ are
\cite{47}:

$A_t = \frac{4}{9} A_s$, $A_u = A_t$, $A_{st} = A_s$, $ A_{su} =
A_{st}$, $A_{tu} = \frac{1}{9} A_s$, $A_s = \frac{8\alpha^2_s}{3}
 \delta_{qq^{'}}$

The differential cross section for the production of gluino pairs
in gluon-gluon collisions is
\begin{eqnarray}
&&\frac{d\sigma}{dt}(gg \rightarrow \tilde{g} \tilde{g}) =  \\ \nonumber
&&\frac{9\pi\alpha^2_s}{4s^2}[
\frac{2(t - m^2_{\tilde{g}})(u - m^2_{\tilde{g}})}{s^2} +
[[\frac{(t-m^2_{\tilde{g}})(u-m^2_{\tilde{g}}) - 2m^2_{\tilde{g}}
(t + m^2_{\tilde{g}})0}{(t-m^2_{\tilde{g}})^2} \\ \nonumber
&&+ \frac{(t-m^2_{\tilde{g}})(u-m^2_{\tilde{g}}) + m^2_{\tilde{g}}
(u - t)}{s(t-m^2_{\tilde{g}})}] + (t \leftrightarrow u) ] \\ \nonumber
&&+\frac{m^2_{\tilde{g}}(s-4m^2_{\tilde{g}})}{(t-m^2_{\tilde{g}})
(u-m^2_{\tilde{g}}}] \,.
\end{eqnarray}
The total cross section has the form
\begin{equation}
 \sigma(gg \rightarrow \tilde{g}\tilde{g}) =
\frac{3\pi\alpha^2_s}{4s}[3[1 + \frac{4m^2_{\tilde{g}}}{s} -
\frac{4m^4_{\tilde{g}}}{s^2}]\ln{[\frac{s+L}{s-L}]}
- [4 + \frac{17m^2_{\tilde{g}}}{s}]\frac{L}{s}] \,,
\end{equation}
where $ L = [s^2 - 4m_{\tilde{g}}^2s]^{1/2}$.

The differential cross section  for the reaction
$q_iq_j \rightarrow \tilde{q}_i \tilde{q}_j$ for the case of
equal masses of righthanded and lefthanded squarks is
\begin{eqnarray}
&&\frac{d\sigma}{dt}(q_iq_j \rightarrow \tilde{q}_i\tilde{q}_j)
= \\ \nonumber
&&\frac{4\pi \alpha^2_s}{9s^2}[- \frac{(t-m^2_i)(t-m^2_j) +st}
{(t- m^2_{\tilde{g}})^2} - \delta_{ij}\frac{(u-m^2_i)(u-m^2_j) +su}
{(u-m^2_{\tilde{g}})^2} +  \\ \nonumber
&&\frac{sm^2_{\tilde{g}}}{(t-m^2_{\tilde{g}})^2} +
\frac{sm^2_{\tilde{g}}}{(u-m^2_{\tilde{g}})^2} \delta_{ij}
- \frac{2sm^2_{\tilde{g}}}{3(t-m^2_{\tilde{g}})(u-m^2_{\tilde{g}})}
\delta_{ij}] \,,
\end{eqnarray}
where $m_i$ and $m_j$ are the masses of produced squarks and
$m_{\tilde{g}}$ is the gluino mass.

For the reaction $q_i\bar{q}_j \rightarrow \tilde{q}_i
\tilde{q}^{*}_j$ the differential cross section has the form
\begin{eqnarray}
&&\frac{d\sigma}{dt}(q_i\bar{q}_j \rightarrow \tilde{q}_i
\tilde{q}^{*}_j) = \\ \nonumber
&&\frac{4\pi\alpha^2_s}{9s^2}[[\frac{ut -m^2_im^2_j}
{s^2}][\delta_{ij}[2 -\frac{2}{3}\frac{s}{(t-m^2_{\tilde{g}})}]
+\frac{s^2}{(t-m^2_{\tilde{g}})^2}] +
\frac{sm^2_{\tilde{g}}}{(t-m^2_{\tilde{g}})^2}] \,.
\end{eqnarray}

For the reaction $gg \rightarrow \tilde{q}_i \tilde{q}_i^{*}$ the
differential cross section is
\begin{eqnarray}
&&\frac{d\sigma}{dt}(gg \rightarrow \tilde{q}_i \tilde{q}_i^{*})
= \\ \nonumber
&&\frac{\pi \alpha^2_s}{s^2}[\frac{7}{48} + \frac{3(u-t)^2}{16s^2}]
[1 + \frac{2m^2t}{(t-m^2)^2} + \frac{2m^2u}{(u-m^2)^2} +
\frac{4m^4}{(t-m^2)(u-m^2)}] \,.
\end{eqnarray}
Here m is the mass of the corresponding squark (we assume the left-
and right-handed squarks are degenerated in mass).

The differential cross section for the reaction
$gq_i \rightarrow gaugino + \tilde{q}_i$ has the form
\begin{eqnarray}
&&\frac{d\sigma}{dt}(gq_i \rightarrow gaugino  + \tilde{q}_i)
= \\ \nonumber
&&\frac{\pi}{s^2}[B_s \frac{(\mu^2 - t)}{s} +
B_t\frac{[(\mu^2 -t)s +2\mu^2(m^2_i-t)]}{(t -\mu^2)^2} + \\ \nonumber
&&B_u\frac{(u-\mu^2)(u+m^2_i)}{(u-m^2_i)^2} +
B_{st}\frac{[(s-m^2_i + \mu^2)(t-m^2_i) -\mu^2s]}{s(t-\mu^2)}
+ \\ \nonumber
&&B_{su}\frac{[s(u  + \mu^2) + 2(m^2_i -\mu^2)(\mu^2 -u)]}
{s(u-m^2_i)} +  \\ \nonumber
&&B_{tu}\frac{[(m^2_i -t)(t +2u +\mu^2) + (t - \mu^2)(s + 2t - 2m^2_i)
+(u - \mu^2)(t + \mu^2 + 2m^2_i)]}{2(t - \mu^2)(u - m^2_i)}] \,,
\end{eqnarray}
where $\mu$ is the mass of the gauge fermion and $m_i$ is the mass of
the scalar quark. The coefficients $B_x$ are contained in
paper of E.Eichten et al.\cite{4}. For instance, for the case when
$gaugino \equiv gluino$
the coefficients $B_x$ are: $B_s = \frac{4\alpha^2_s}{9}
\delta_{ij}$, $B_t = \frac{9}{4}B_s$, $B_u = B_s$, $B_{st} = -B_t$,
$B_{su} = \frac{1}{8} B_s$, $B_{tu} = \frac{9}{8}B_s$.

 Consider finally the production of sleptons. The differential cross
section for the production of charged slepton-sneutrino pairs is
\begin{equation}
\frac{d\sigma}{dt}(d\bar{u}  \rightarrow W \rightarrow \tilde{l}_L
\bar{\tilde{\nu}}_L) = \frac{g^4_2|D_W(s)|^2}{192\pi s^2}
(tu - m^2_{\tilde{l}_L}m^2_{\tilde{\nu}_L}) \,.
\end{equation}
For $\tilde{l}_L$ pair production the differential cross section is
\begin{eqnarray}
&&\frac{d\sigma}{dt}(q\bar{q} \rightarrow \gamma^{*}, Z \rightarrow
\tilde{l}_L \bar{\tilde{l}}_L) =
\frac{2\pi \alpha^2}{3s^2}[tu - m^4_{\tilde{l}_L}]
[\frac{q^2_l q^2_q}{s^2} + \\ \nonumber
&&(\alpha_l
- \beta_l)^2(\alpha^2_q + \beta^2_q)|D_Z(s)|^2 + \\ \nonumber
&&\frac{2q_lq_q\alpha_q(\alpha_l - \beta_l)(s  - M^2_Z)}{s}
|D_Z(s)|^2] \,,
\end{eqnarray}
where $D_V(s) = 1/(s - M^2_V + iM_V \Gamma_V)$, $q_l = -1$, $q_{\nu}
= 0$, $q_u = 2/3$, $q_d = -1/3$, $\alpha_l = \frac{1}{4}(3t - c)$,
$\alpha_{\nu} = \frac{1}{4}(c + t)$, $\alpha_u = -\frac{5}{12}
t + \frac{1}{4}c$, $\alpha_{d} = -\frac{1}{4}c +\frac{1}{12}t$,
$\beta_l =\frac{1}{4}(c + t)$, $\beta_{\nu} = -\frac{1}{4}(c + t)$,
$\beta_{u} = -\frac{1}{4}(c + t)$, $\beta_{d} = \frac{1}{4}(c + t)$,
$c = \cot{\theta_W}$, $t = \tan{\theta_W}$.
The differential cross section for sneutrino pair production can
be obtained by the replacement $\alpha_l$, $\beta_l$, $q_l$ and
$m_{\tilde{l}}$ by $\alpha_{\nu}$, $\beta_{\nu}$, $0$ and $m_{\tilde{\nu}}$
respectively, whereas for $\tilde{l}_R$ pair production one has
substitute $\alpha_l - \beta_l \rightarrow \alpha_l + \beta_l$ and
$m_{\tilde{l}_L} \rightarrow m_{\tilde{l}_R}$.

\subparagraph{Superparticle decays.}

The decay widths of the superparticles depend rather
strongly on the relations between superparticle masses.
Here we outline the main decay channels only.
The formulae for the decay widths are contained in refs. \cite{49}.
Consider at first the decays of gluino and squarks. For
$m_{\tilde{g}} > m_{\tilde{q}}$ the main decays
are the following:
\begin{equation}
\tilde{g} \rightarrow \tilde{q}_i \bar{q}_i\,, \bar{\tilde{q}}_i q_i \,,
\end{equation}
\begin{equation}
\tilde{q}_k \rightarrow \chi^{0}_{i} q_k \,,
\end{equation}
\begin{equation}
\tilde{q}_k \rightarrow \chi^{+}_j q_m,\, \chi^{-}_j q_l \,,
\end{equation}
For $m_{\tilde{g}} < m_{\tilde{q}}$ the main decays are:
\begin{equation}
\tilde{q}_i \rightarrow \tilde{g} q_i \,,
\end{equation}
\begin{equation}
\tilde{g} \rightarrow q \bar{q}^{'} \chi^{+}_{k} \,,
\end{equation}
\begin{equation}
\tilde{g} \rightarrow q^{'} \bar{q} \chi^{-}_{k} \,,
\end{equation}
\begin{equation}
\tilde{g} \rightarrow q \bar{q} \chi^{0}_{k} \,.
\end{equation}
The charginos and neutralinos usually are supposed to be lighter than
gluino and squarks and their main decays are:
\begin{equation}
\chi_{i}^{0}  \rightarrow \chi_{j}^{0} + Z \,,
\end{equation}
\begin{equation}
\chi^0_i \rightarrow \chi_{j}^{\pm} + W^{\mp} \,,
\end{equation}
\begin{equation}
\chi^{\pm}_i \rightarrow \chi^0_j + W^{\pm} \,,
\end{equation}
\begin{equation}
\chi^{\pm}_i \rightarrow  \chi^{\pm}_j + Z  \,.
\end{equation}
Two-body decays of neutralinos and charginos
into Higgs bosons are:
\begin{equation}
\chi^0_i \rightarrow \chi^0_j +h(H) \,,
\end{equation}
\begin{equation}
\chi^0_i \rightarrow \chi^{\pm}_k + H^{\mp} \,,
\end{equation}
\begin{equation}
\chi^{\pm}_i \rightarrow   \chi^0_k + H^{\pm} \,,
\end{equation}
\begin{equation}
\chi^{\pm}_i \rightarrow \chi^{\pm}_j + h(H)   \,.
\end{equation}

The left sleptons dominantly decay via gauge interactions
into charginos or neutralinos via two body decays
\begin{equation}
\tilde{l}_L \rightarrow l + \chi_i^0 \,,
\end{equation}
\begin{equation}
\tilde{l}_L \rightarrow \nu_{L} +\chi^{-}_j \,,
\end{equation}
\begin{equation}
\tilde{\nu}_L \rightarrow \nu_{L} + \chi_{i}^0 \,,
\end{equation}
\begin{equation}
\tilde{\nu}_L \rightarrow l + \tilde{\chi}^{+}_j  \,.
\end{equation}
For relatively light sleptons only the decays into the LSP are
possible, so that light sneutrino decays are invisible. Heavier
sleptons can decay via the chargino or other (non LSP)
channels. These decays are important because they proceed
via the larger $SU(2)$ gauge coupling constant and can dominate
the direct decay to LSP. The $SU(2)$ singlet charged sleptons
$\tilde{l}_R$ only decay via their $U(1)$ gauge interactions
and in the limit of vanishing Yukawa coupling their decays to
charginos are forbidden. Therefore the main decay mode of
righthanded slepton is
\begin{equation}
\tilde{l}_R \rightarrow l + \chi^{0}_{i} \,.
\end{equation}
In many cases the mode into LSP dominates.

Let us now briefly describe the main signatures for the search for
sparticles at LHC. Sparticle pair production at LHC is followed
by sparticle decays untill the LSP is reached. Therefore, the main
signature of sparticle production are the events with
$(n \geq 0)$ jets plus $m \geq 0$ isolated leptons plus missing
transverse energy due to escaping from detector registration 2 LSP.
It is natural to divide the signatures into the
following categories \cite{50}:

a. multi jets plus $E^{miss}_t$    events,

b. 1l  plus jets plus $E_t^{miss}$ events,

c. 2l plus jets plus $E_t^{miss}$ events,

d. 3l plus jets plus $E_t^{miss}$ events,

e. 4l plus jets plus $E_t^{miss}$ events,

f. $\geq 5 l$  plus jets plus $E_t^{miss}$ events.

Multileptons arise as a result of the cascade decays of neutralinos
and charginos into W- and Z-bosons with subsequent decays of
W- and Z-bosons into leptonic modes. For instance, the same sign
dilepton events  arise as a result of the cascade decay
\begin{equation}
\tilde{g} \rightarrow q^{'} \bar{q} \chi^{\pm}, \; \chi^{\pm}
\rightarrow W^{\pm}\chi^{0}_1  \rightarrow l^{\pm} \nu \chi^0_1 \,,
\end{equation}
where l stands for both $e$ and $\mu$.
Opposite sign dilepton events can arise as a result of cascade
decay
\begin{equation}
\tilde{g} \rightarrow q \bar{q} \chi^{0}_{i} , \;
\chi^0_i \rightarrow Z \chi^0_1 \rightarrow
l^{+}l^{-} \chi^0_1 \,.
\end{equation}
It should be noted that multilepton supersymmetry
signatures arise as a result of decays
of squarks or gluino into charginos or neutralinos different from
LSP with subsequent decays of charginos or neutralinos into (W,
Z)-bosons plus LSP. Leptonic decays of (W, Z)-bosons is the
origin of leptons.  However, for the case of nonuniversal gaugino
mass relations at GUT scale it is possible to realize the
situation \cite{42} when all charginos and neutralinos are heavier
than gluino and squarks. Therefore, gluino and squarks will
decay mainly into quarks or gluons plus LSP so cascade decays and as a
consequence multilepton events will be absent.

\paragraph{3.3.2  The search for SUSY Higgs bosons}

\subparagraph{  }

In MSSM there are four Higgs bosons
$(h,H,A,H^{\pm})$. As it has been mentioned before at tree level
the lightest Higgs boson mass is predicted to be lighter than
$m_{Z}$. However an account of radiative corrections for big top
quark mass (that takes place in reality) can increase the
Higgs boson mass up to 120 Gev for stop mass equal to 1 Tev.
From the vacuum stability bound \cite{15} the standard Higgs boson
mass for $m_{t} \ge 175 Gev$ is predicted to be heavier than 120 Gev.
It means that the predictions for the Higgs boson mass
in MSSM and SM lie in different mass intervals, i.e. the knowledge
of the Higgs boson mass allows to discriminate between MSSM and SM.
In particular, the discovery of
light Higgs boson at LEP2 will be powerful untrivial evidence in
favour of low energy broken supersymmetry.

\subparagraph{$h, H \rightarrow \gamma\gamma$.}

In the MSSM, the neutral Higgs boson remains extremely narrow in the
kinematic region where the two photon decay has a reasonable
branching ratio. The experimental requirements and backgrounds
are the same as in the case of standard Higgs boson decaying
into two photons. Therefore the ECAL mass resolution and
acceptance are crusial for the Higgs boson discovery. For the case
when SUSY masses are bigger than $O(300)$ Gev the branching of
$h \rightarrow \gamma\gamma$ in MSSM  coincides with the
corresponding branching of SM, so we have in fact the same
significance as in the case of SM. We shall consider the limiting
case when the sparticle masses are heavy enough and they don't
play important role in Higgs boson  decays. So the  decay channels
for the MSSM lightest Higgs boson are the same as for the SM Higgs
boson, but the production rates are significantly modified by
MSSM couplings. There are two main production processes for MSSM
neutral Higgs bosons: $gg \rightarrow h(H)$ and $gg \rightarrow h(H)
b\bar{b}$. The second mechanism is important at large $\tan{\beta}$
due to enhanced $h(H)b\bar{b}$ couplings. These associated production
processes are interesting, as b-tagging techniques may be able to
enhance the signal/background ratio, possibly allowing the observation
of h, H and A \cite{51}.

\subparagraph{$H \rightarrow ZZ^{*}, ZZ$ and $h \rightarrow ZZ^{*}$.}

The scalar Higgs bosons H and h couple to W and Z boson pairs, and so
may be searched for in 4-lepton final states from $h,H \rightarrow
ZZ^{*}, ZZ$. The regions of MSSM $(\tan(\beta), m_{A})$-space
($\tan(\beta) = \frac{<H_{t}>}{<H_{b}>}$ ,$m_{A}$ is the mass of
the axial Higgs boson), where the Higgs bosons could be discovered
through four lepton modes are divided into the region of low
$\tan({\beta})$ where H could be discovered and the high
$\tan({\beta})$ region where only h could be discovered.

\subparagraph{$h, H, A \rightarrow \tau\bar{\tau} \rightarrow
l^{\pm}h^{\pm} + X$.}

The $\tau \bar{\tau}$ final states can be searched for in a 'lepton +
hadron' final state or in a $e + \mu$ final state.
For the one lepton plus one hadron final states, intermediate
backgrounds are due to $Z, \gamma^{*} \rightarrow \tau \bar{\tau};
t\bar{t} \rightarrow \tau \bar{\tau} + X, \tau + X$ and
$b\bar{b} \rightarrow \tau \bar{\tau} + X, \tau X$. Reducible backgrounds
are due to the events with one hard lepton and jets with a jet
misidentified as a $\tau$.
Typically cuts are the following \cite{1}:

1. One isolated lepton with $p_{T} \ge (15 - 40)Gev$ depending on
the axial Higgs boson mass $m_{A}$ and $|\eta| \le 2$.

2. One $\tau$-jet candidate which contains only one charged hadron.

3. No other significant jet activity.

An overall lepton reconstruction of 90 percent is assumed. For the
$\tau \bar{\tau} \rightarrow $ $ e^{\pm} + \mu^{\mp} $
final states a pair of opposite
sign isolated electrons and muons with $p_{T} \ge 20 $ Gev and $|\eta|
\le 2$ is required. There can be large backgrounds from $Z,
\gamma^{*} \rightarrow \tau\bar{\tau}, t\bar{t}, b\bar{b}$ and
$W^{+}W^{-}$. The $t\bar{t}$ and $W^{+}W^{-}$ backgrounds can be
reduced to the level of $\approx 20$ percent of the $Z,\gamma^{*}$
backgrounds. Roughly speaking $A,H$ bosons can be discovered using
these modes with the masses up to 600 Gev.

\subparagraph{Charged Higgs $H^{\pm}$ in $ t \rightarrow
H^{\pm}h$, $H^{\pm} \rightarrow \tau \nu_{\tau}$.}

In the MSSM the top quark can decay to a charged Higgs($t \rightarrow
H^{+}b$). The $t \rightarrow H^{+}b$ branching ratio is large at low
and large $\tan(\beta)$ values, having a minimum at $\tan(\beta)
\approx 6$. The $H^{+}$ has two main decay modes, $H^{+} \rightarrow
c\bar{s}$ and $H^{+} \rightarrow \tau^{+} \nu_{\tau}$. The
$H^{+} \rightarrow \tau^{+} \nu_{\tau}$ branching is large for
$\tan(\beta) \ge 2$, and only slightly depends on $\tan(\beta)$.

\subparagraph{$h,H,A \rightarrow \mu^{+}\mu^{-}$.}

In the SM and in the MSSM for interesting values of the Higgs boson
masses the branching ratio of $H \rightarrow \mu^{+}\mu^{-}$ is
small $ \approx 3\cdot10^{-4}$. For the SM Higgs boson
the main background is the Drell-Yan production
$\gamma{*},Z \rightarrow \mu^{+}\mu^{-}$. However in the MSSM
the $\mu^{+}\mu^{-}$ channel could be very interesting for large
values of $\tan(\beta)$. It appears that for $m_{A} \le 200$
Gev and $\tan(\beta) \ge 2.5$ for integrated luminosity
$10^{4}pb^{-1}$ it is possible to discover A and H bosons at the
level $\ge 7\sigma$. The discovery potential of the Higgs boson
for this mode at $L = 10^{5}$ $pb^{-1}$  is similar to the
discovery potential of the Higgs boson for the standard mode
$H,h,A \rightarrow \tau\tau$ for $10^{4}pb^{-1}$. However the
$\mu^{+}\mu^{-}$ channel gives a much better signal identification and
mass resolution \cite{1}.

\subparagraph{h,H,A in associated production $\bar{b}bH_{susy}$.}

Whilst the $gg \rightarrow b\bar{b}H$ associated production is
negligible for the SM Higgs boson compared to $t\bar{t}H$ in the
MSSM the rate of $gg \rightarrow b\bar{b}H_{susy}$ is (30 - 50)
percent of the total production for $m_{A} = 100$ Gev and (70 - 80)
percent for $m_{A} = 300$ Gev with $\tan(\beta) = 10 - 30$. The
signal-to-background ratio can be enhanced by b-tagging.

\subparagraph{$A \rightarrow Zh \rightarrow b\bar{b}b\bar{b}$.}

The large branching ratio ($\approx 50$ percent) of $A \rightarrow
Zh$ in the region $\tan(\beta) \le 2$ for 180 Gev$ \le m_{A} \le
2m_{t}$ could allow the observation of $A$ and $h$ in this region.

The main conclusion concerning the situation with the search for
MSSM Higgs bosons for different $(m_{A},\tan(\beta))$ values is that
Higgs bosons for $m_{A} \le 500$ Gev would be detectable at CMC
except may be very difficult region for 110 Gev$ \le m_{A} \le $200
Gev, $3 \le \tan(\beta) \le 10$ and a smaller one for
$\tan(\beta) \approx 2.5$ and 200 Gev $ \le m_{A} \le $ 280 Gev.
The most promising channel for these regions is $h,H \rightarrow
\bar{b}b$, from $Wh, Zh$ and $t\bar{t}h$ final states.

\paragraph{3.3.3 Squark and gluino search}

\subparagraph{Jets + $E^{miss}_{T}$ channel.}

There are a lot of possible scenario which depend on the concrete
values of squark and gluino masses for the search for
squarks and gluino using this signature. Consider two
typical scenario \cite{1}.

Scenario A. : $m_{\tilde{b}}, m_{\tilde{t}} \le m_{\tilde{g}}$ .

In this case the two body decay $\tilde{g} \rightarrow
\tilde{b}\bar{b}$ with $\tilde{b} \rightarrow \chi^{0}_{1}b$
dominates, because the top is too massive to allow
$\tilde{g} \rightarrow \bar{t}\tilde{t}$ decay. Thus the resulting
event signature will have 4 b-jets in final state.

Scenario B. : $m_{\tilde{q}} = m_{\tilde{g}}$.

For such scenario the gluino decays directly into
LSP $(\tilde{g} \rightarrow q\bar{q}\chi^0_1)$ leading to the
typical $E_{T}^{miss} + multijets$ final state event topology
without an excess of b-jets.

In scenario with heavy squarks and gluino($m_{\tilde{q}} =1550$ Gev,
$m_{\tilde{g}} = 1500$ Gev) after using some cuts we have
$\frac{N_{ev}}{N_{back}} \sim 8.5$. In general for an integrated
luminosity of $10^{5}(pb^{-1} - 10^{3}pb^{-1})$, the sensitivity to
the channel of multijets + missing energy can be expressed as a
mass reach for gluino mass $m_{\tilde{g}}$ \cite{2}:

$m_{\tilde{g}} \le 1600(1050) Gev$ for $m_{\tilde{q}} =
2m_{\tilde{g}}$ ,

$m_{\tilde{g}} \le 2300(1800) Gev$ for $m_{\tilde{q}} =
m_{\tilde{g}}$ ,

$m_{\tilde{g}} \le 3600(2600) Gev$ for $m_{\tilde{q}} =
\frac{m_{\tilde{g}}}{2}$ .

\subparagraph{Lepton(s) + $E^{miss}_t$ + jets.}

Cascade decays of squarks and gluino are an important source of leptons.
Therefore, events with $E^{miss}_t$ + jets + lepton(s) provide a good
signature  for  searching for gluinos and squarks over a wide mass range.
Four event classes, containing one to three leptons(muons or electrons)
give important signatures for the search for squarks and gluino:

$1l$ : a single lepton + $E^{miss}_t$ + jets,

$2l$: two leptons with opposite charges + $E_t^{miss}$ + jets,

$2l(ss)$: same-sign dileptons + $E^{miss}_t$ + jets,

$3l$: three dileptons + $E_t^{miss}$ + jets.

The  background comes from $t\bar{t}$, $W$ + jets, $Z$ + jets,
$WW$, $WZ$. The main conclusion \cite{52} is that the signature
with a single lepton + $E^{miss}_t$ gives the most powerful restriction
on the gluino mass for the search for SUSY at CMS detector. Namely, for
$L_t$ from $10^{4}$ $pb^{-1}$ to $10^{5}$ $pb^{-1}$  the CMS will be
able to discover gluino from 1.5 Tev up to 2 Tev and squarks
from 1.5 Tev up to 2.3 Tev. Other conclusion is that the gluino and
squark discovery mass limits are not very sensitive to increase of
SM background.

\paragraph{3.3.4 Neutralino and chargino search}

\subparagraph{ }

Chargino and neutralino pairs, produced through the Drell-Yan mechanism
or squark exchange may be detected through their leptonic decays
$\chi_{1}^{\pm}, \chi^0_2 \rightarrow lll + E^{miss}_T$.
The leptonic decays of $\chi^{\pm}_1$ and $\chi^0_2$ are the following:

$\chi^{\pm}_1 \rightarrow \chi^0_1 l^{\pm} \nu$ ,

$\chi^{\pm}_1 \rightarrow (\tilde{l}^{\pm}_{L,R} \rightarrow
\chi^0_1l^{\pm}) \nu$ ,

$\chi^{\pm}_1 \rightarrow (\tilde{\nu}_L \rightarrow \chi^0_1
\nu)l^{\pm}$ ,

$\chi^{\pm}_1 \rightarrow (W^{\pm} \rightarrow l^{\pm} \nu) \chi^0_1$ ,

$\chi^0_2 \rightarrow \chi^0_1 l^+l^-$ ,

$\chi^0_2 \rightarrow (\tilde{l}^{\pm}_{L,R} \rightarrow
\chi^0_1 l^{\pm}) l^{\mp}$ ,

$\chi^0_2 \rightarrow (\chi^{\pm}_1 \rightarrow \chi^0_1 l^{\pm} \nu)
l^{\mp} \nu$ .

The three-lepton
signal is produced through the decay chain $\chi^{\pm}_1 \rightarrow
l^{\pm} + \chi^0_1$ and $\chi^0_2 \rightarrow ll + \chi^0_1$ ,
where the undetected neutrino and $\chi^0_1$ produce $E^{miss}_{T}$.
The main backgrounds to this channel arise from $WZ/ZZ$,
$t\bar{t}$, $Zb\bar{b}$ and $b\bar{b}$  production.
In principle there could be SUSY background arising as a result of squark
and gluino cascade decays into multileptonic modes.

Typical cuts are the following \cite{53}:

i. Three isolated leptons with $p^l_t > 15$ Gev.

ii. Veto central jets with $E_t > 25$ Gev in $|\eta| < 3.5$.

iii. $m_{l\bar{l}} < 81$ Gev or $m_{l\bar{l}} \neq M_Z \pm
\delta M_Z$.

The main conclusion is that neutralino and chargino could be detected
provided their masses are lighter than 350 Gev \cite{53}. Moreover, it is
possible to determine the $M(\chi^0_2) - M(\chi^0_1)$ mass difference by
the measurement of the distribution on $l^+l^-$ invariant mass arising
as a result of the decay $\chi^0_2 \rightarrow \chi^0_1 + l^+l^-$
\cite{53}.

\paragraph{3.3.5 Sleptons search}

\subparagraph{ }

Slepton pairs, produced through the Drell-Yan mechanism can be
detected through their leptonic decays $\tilde{l} \rightarrow l +
\chi^0_1$. So in the final state we expect dilepton pair with
missing energy and no hadronic jets. Here we shall use the
results of ref. \cite{54} where concrete estimates have been made for
CMS detector. Namely, we consider two points of ref. \cite{54}

Point A: $m(\tilde{l}_L) = 314$ Gev, $m(\tilde{l}_R) = 192$ Gev,
$m(\tilde{\nu}) = 308$ Gev, $m(\tilde{\chi}^0_1) = 181$ Gev,
$m(\tilde{\chi}^0_2) = 358$ Gev, $m(\tilde{g}) = 1036$ Gev,
$m(\tilde{q}) = 905$ Gev, $\tan(\beta) = 2$, $sign (\mu) = -$.

Point B: $m(\tilde{l}_L) = 112$ Gev, $m(\tilde{l}_R) = 98$ Gev,
$m(\tilde{\nu}) = 93$ Gev, $m(\tilde{\chi}^0_1) = 39$ Gev,
$m(\tilde{\chi}^0_2) = 87$ Gev, $m(\tilde{g}) = 254$ Gev,
$m(\tilde{q}) = 234$ Gev, $\tan{\beta} = 2$, $sign{\mu} = -$ .

For point A the following cuts have been used: $p^l_t \geq 50$ Gev,
$Isol \leq 0.1$, $|\eta| \leq 2.5$, $E_t^{miss} \geq 120$ Gev,
$\Delta \phi(E^{miss}_t, ll) \geq 150^o$, jet veto - no jets with
$E_t^{jet} \geq 30$ Gev in $|\eta| \leq 4.5$, Z-mass cut -
$(M_Z \pm 5$ Gev excluded), $\Delta\phi(l^+l^-) \leq 130^o$.

With such cuts for the total luminosity $L_t = 10^5 pb^{-1}$ 91
events with $e^+e^- + \mu^+ \mu^-$ resulting from slepton decays
have been found. The standard WS model background comes from $WW$,
$t\bar{t}$, $Wt\bar{b}$, $WZ$ $\tau \bar{\tau}$ and gives 105 events.
No SUSY background have been found. The significance
$ S = \frac{Sleptons}{\sqrt{Background + Sleptons}}$ for the
slepton discovery at the point A is $S =6.5 $.

For the point B the cuts are similar to the point A,
except $p^t_l \geq 20$ Gev, $E^{miss}_t \geq 50$ Gev,
$\Delta \phi (E^{miss}_t,ll)
\geq 160^{o}$. For the total luminosity
$L_t = 10^{4}pb^{-1}$ the
number of $e^{+}e^- + \mu^{+} \mu^{-} $ events resulting from direct
slepton production has been found to be 323. The number of the
background events have been estimated equal to 989(standard
model background) + 108(SUSY background) =
1092. The significance is equal to $S = 8.6$.

The main conclusion of the ref. \cite{54}  is
that for $L_{t} = 10^{5}pb^{-1}$ CMS will be able to discover
sleptons with the masses up to 400 Gev.

\subparagraph{The search for flavour lepton number violation in
slepton decays.}

In supersymmetric models with explicit flavour lepton number violation
due to soft supersymmetry breaking terms there could be detectable
flavour lepton number violation in slepton decays \cite{55}.
For instance, for the case of nonzero mixing $\sin{\phi} \neq 0$ between
righthanded selectrons and smuons we have flavour lepton number
violation in slepton decays, namely \cite{55}:
\begin{equation}
\Gamma(\tilde{\mu}_R \rightarrow \mu + LSP) =
\Gamma \cos^2{\phi}\,,
\end{equation}
\begin{equation}
\Gamma(\tilde{\mu}_R \rightarrow e + LSP) = \Gamma \sin^2{\phi}\,,
\end{equation}
\begin{equation}
\Gamma(\tilde{e}_R \rightarrow e + LSP) = \Gamma \cos^2{\phi}\,,
\end{equation}
\begin{equation}
\Gamma(\tilde{e}_R \rightarrow \mu + LSP) = \Gamma \sin^2{\phi}\,,
\end{equation}
\begin{equation}
\Gamma = \frac{g^2_1}{8\pi}(1 - \frac{M^2_{LSP}}{M^2_{SL}})^2 \,.
\end{equation}

The typical prediction of the nonzero smuon-selectron mixing is
the existence of accoplanar $e^{\pm}\mu^{\mp}$ signal events with missing
energy arising as a result of the production of slepton pairs with their
subsequent decays with flavour lepton number violation. The possibility
to detect flavour lepton number violation in slepton decays at LHC
has been discussed in ref. \cite{56}. The main conclusion is that for
the most optimistic case of the maximal mixing $\sin{\phi} = \frac{1}
{\sqrt{2}}$ it would be possible to discover slepton mixing at LHC
for the points A and B which have been considered for the case of
zero slepton mixing in ref. \cite{54}.

\subsection{The search for physics beyond SM and MSSM}

\paragraph{Search for new vector bosons.}

Many string inspired supersymmetric electroweak models and grand
unified models based on extended gauge groups ($SO(10)$, $E_6$...)
predict the existence of new relatively light neutral $Z^{'}$ bosons.
The main mechanism for the production of such new neutral vector
bosons is the quark-antiquark fusion. The cross section is
given by the standard formula:
\begin{eqnarray}
&&\sigma (pp \rightarrow Z^{'} + ...) = \sum_{i}
\frac{12\pi^{2}\Gamma(Z^{'} \rightarrow \bar{q}_iq_i)}{9M_{Z^{'}}s}
\int_{M^2_{Z^{'}}/s}^{1} \frac{dx}{x} \\ \nonumber
&&[\bar{q}_{pi}(x,\mu)
q_{pi}(x^{-1}M^2_{Z^{'}}s^{-1}, \mu)
 +q_{pi}(x,\mu)\bar{q}_{pi}(x^{-1}M^2_{Z^{'}}s^{-1}, \mu) \,.
\end{eqnarray}
Here $\bar{q}_{pi}(x, \mu )$ and $q_{pi}(x,\mu )$
are the parton distributions of the antiquark $\bar{q}_i$ and
quark $q_i$ in the proton at the normalization
point $\mu \sim M_{Z^{'}}$ and
$\Gamma(Z^{'} \rightarrow \bar{q}_{i}q_i)$ is the hadronic decay width of
the $Z^{'}$ boson into quark-antiquark pair with a flavor i.
In most models  as a consequence of the $\gamma_{5}$ anomaly cancelation
$Z^{'}$ boson interacts both with quarks and leptons, therefore
the best signature for the search for $Z^{'}$ boson is through
its decay to electron pairs, muon pairs and jet pairs.
The LHC $Z^{'}$ boson discovery potential depends on the
couplings of $Z^{'}$ boson with quarks and leptons.
For the $Z^{'}$ boson decaying into lepton pair the main
background comes from the Drell-Yan process which is under control.
For $Z^{'}$ with  quark and lepton couplings equal to Z-couplings
with quarks and leptons it would be possible to discover the
$Z^{'}$-boson with a mass up to 5 Tev \cite{2}.

Many extended gauge electroweak models based for instance on
the gauge group $SU(2)_L \otimes SU(2)_R \otimes U(1)$ predict
the existence of the additional charged vector $W^{'}$ boson.
The main production mechanism for the $W^{'}$ boson is the
quark-antiquark fusion similar to the case of $Z^{'}$
production. The best way to look for $W^{'}$ boson is through
its leptonic mode $W^{'} \rightarrow l \nu$.

For the model with righthanded charged $W^{'}$ boson it would be
possible to discover the  $W^{'}$ boson through leptonic mode
$W^{'} \rightarrow e\nu$ with $W^{'}$ mass up to 6 Tev \cite{2}.
Typical accuracy in the determination of the
$W^{'}$ mass is (50 -100) Gev.

\paragraph{Search for supesymmetry with R-parity violation.}

Most of supersymmetric phenomenology assumes the MSSM which
conserves R-parity. As a consequence of R-parity conservation
supersymmetric particles can only be produced in pairs and a
supersymmetric state cannot decay into conventional states.
This has untrivial consequence for the search for
supersymmetric particles at supercolliders; in particular
all experimental searches of SUSY rely on pair production and
on missing transverse momentum $p_t^{miss}$ as a signal for the
production of the LSP, which must be stable and electrically neutral.
However, at present there are no deep theoretical
motivations in favour of R-parity conservation.  The phenomenology
of the models with explicit R-violation at hadron colliders
has been studied in refs. \cite{57}.
 The most general trilinear terms in superpotential explicitly violating
R-parity have the form \cite{57}
\begin{equation}
W_{R,br} = \lambda_{ijk} L_i L_j\bar{E}_k
+\lambda^{'}_{ijk}L_iQ_j\tilde{D}_k + \lambda^{''}_{ijk}\bar{U}_i
\bar{D}_j \bar{D}_k\,,
\end{equation}
where $L$ and $\bar{E}$ ($Q$ and $\bar{U}$, $\bar{D}$) are the
(left-handed) lepton doublet and the antilepton singlet (quark
doublet and antiquark singlets) chiral superfields respectively.
The terms of (233) violate baryon and lepton number and, if present
in the lagrangian, they generate an unacceptably large amplitude
for proton decay suppressed only by the inverse squark mass squared.
The R-parity prohibits the dangerous terms (233) in the
superpotential. However, R-parity is not the single way to construct
a minimal supersymmetric extension of the standard model.  It is easy to
write down alternative to R-parity symmetries which allow for a
different set of couplings. For example, under the transformation
\begin{equation}
(Q,\bar{U}, \bar{D}) \rightarrow -(Q,\bar{U}, \bar{D}),\;\;
(L,\bar{E}, H_{1,2}) \rightarrow +(L,\bar{E}, H_{1,2}) \,,
\end{equation}
only the quark superfields change sign. If the lagrangian is invariant
under transformations (234) then only the last, baryon number
violating term in (233)
$\bar{U} \bar{D} \bar{D}$ is forbidden. This gives a new model in which
a single supersymmetric state can couple to standard model states
breaking R-parity. Similarly, there are analogous transformations
forbiding the lepton number violating terms.

In the direct search for supersymmetric particles the phenomenology
is altered considerably when including R-parity violating terms in the
superpotential. In general both the production mechanisms and the
decay patterns  can change. Other than the standard supersymmetric pair
production of particles there is now the possibility of production
of R-odd final states as well. Also, if all supersymmetric particles
decay in the detector, we will no longer have the standard $p_t^{miss}$
signal and the decay patterns will all be altered. In particular, the LSP
will decay mainly into three-body final states \cite {57}.
However, except for LSP, which now decays, all particles
predominantly decay as in the MSSM.
Consider the case when LSP decays within the detector. For the model
with R-parity breaking terms involving leptonic
superfields we expect additional lepton pairs in the final state
as a result of the LSP decay. Therefore we expect instead of missing energy
signature the presence of additional lepton pairs compared to
standard MSSM signatures that is in principle more visible at LHC
than the SUSY signatures in MSSM. The $\bar{U} \bar{D} \bar{D}$
operators however lead to less characteristic signals, but for
cascade decays they lead to signals compartible to those in the
MSSM.

Note that it is possible to construct the model with
supersmall R-parity violation and with relatively
longlived $t \sim (10^{-1} - 10^{-9})$ sec
charged $\tilde{\tau}_R$ slepton playing the role of LSP \cite{58}.
Relatively longlived charged LSP penetrates through the detector
and it is possible to detect it by the track measurement.

\paragraph{Search for leptoquarks.}

At LHC the leptoquark pair production proceeds dominantly through
gg fusion, which does not involve any lepton-quark-leptoquark
vertex and therefore is predicted with $\approx 50$ percent accuracy.
In 25 percent of the events, the final state contains 2 electrons
and 2 jets. The dominant background in this case is from $t\bar{t}$
production.  The conclusion is that leptoquarks with masses up to
1 Tev will be discovered at LHC.

\paragraph{Search for scalar colour octets.}

Relatively light ($M_8 \leq O(1)$ Tev)
scalar $SU_c(3)$ colour
$SU_L(2)\otimes U(1)$ neutral octets are predicted in
some supersymmetric and nonsupersymmetric GUTs \cite{60}. Light scalar
octets naturally arise in models with big compactification radius
of the additional space dimensions.

To be precise, consider light scalar octets neutral under $SU_L(2)
\otimes U(1)$ electroweak gauge group. Such particles are described
by the selfconjugate scalar field $\Phi^{\alpha}_{\beta}(x)$
($(\Phi^{\alpha}_{\beta}(x))^{*} = \Phi^{\beta}_{\alpha}(x),
\sum_{\alpha}\Phi^{\alpha}_{\alpha}(x) =0$) and they
interact only with gluons. Here $\alpha= 1,2,3;\; \beta= 1,2,3$ are the
$SU(3$) indices. The scalar potential for the scalar octet field
$\Phi^{\alpha}_{\beta}(x)$ has the form
\begin{equation}
V(\Phi) = \frac{M^2}{2}Tr(\Phi^{2}) + \frac{\lambda_{1}M}{6}Tr(\Phi^{3})
+ \frac{\lambda_{2}}{12}Tr(\Phi^{4})
+ \frac{\lambda_{3}}{12}(Tr(\Phi^{2})^2 \,.
\end{equation}
The term $\frac{\lambda_{1}M}{6}Tr(\Phi^{3})$ in the scalar
potential (235) breaks the
discrete symmetry $\Phi \rightarrow -\Phi$. The existence
of such term in the Lagrangian leads to the decay of scalar octet mainly
into two gluons through one-loop diagrams similar to the corresponding
one-loop diagrams describing the Higgs boson decay into two photons.
One can find that the decay width of the scalar octet into two
gluons is determined by the formula \cite{59}
\begin{equation}
\Gamma(\Phi \rightarrow gg) = \frac{15}{4096\pi^{3}}\alpha_{s}^{2}c^2
\lambda^{2}_{1}M \,,
\end{equation}
where
\begin{equation}
c = \int_{0}^{1}\int^{1-w}_{0}du\,dw\, \frac{wu}{1-u-w} \simeq 0.48
\end{equation}
and $\alpha_{s}$ is the effective coupling constant at some
normalization point $\mu \sim M$. Numerically for $\alpha_{s} = 0.12$
we find that
\begin{equation}
\Gamma(\Phi \rightarrow gg) = 0.4\cdot 10^{-8}\lambda_{1}^{2}M \,.
\end{equation}
From the requirements that colour $SU(3)$ symmetry is unbroken
(the minimum $<\Phi^{\alpha}_{\beta}(x)> = 0$ is the deepest one)
and the effective coupling constants $\bar{\lambda}_{2}$,
$\bar{\lambda}_{3}$ don't have Landau pole singularity up
to the energy $M_o = 100M$ we find that $\lambda_{1} \leq  O(1)$.
Therefore the decay width of the scalar colour octet is less
than $O(1)Kev$, $O(10)Kev$ for the octet masses 100, 1000 Gev
respectively. It means that new hadrons composed from scalar octet
$\Phi$, quarks and gluons ($g\Phi,\, \bar{q} \Phi q, \, qqq \Phi$)
are relatively longlived even for high scalar octet mass.
Consider the pair production of scalar octets at LHC. The corresponding
lowest order parton cross sections have the form \cite{59}
\begin{equation}
\frac{d\sigma}{dt}(\bar{q}q \rightarrow   \Phi \Phi) =
\frac{4\pi\alpha^{2}_{s}}{s^4}(tu - M^4) \,,
\end{equation}
\begin{eqnarray}
&&\frac{d\sigma}{dt} = \frac{\pi \alpha^{2}_s}{s^2}(\frac{7}{96} +
\frac{3(u-t)^2}{32s^2})(1 + \frac{2M^2}{u-M^2} + \frac{2M^2}{t-M^2}
+ \frac{2M^4}{(u-M^2)^2} + \\ \nonumber
&&\frac{2M^4}{(t-M^2)^2} +
\frac{4M^4}{(t-M^2)(u-M^2)}) \,,
\end{eqnarray}
\begin{equation}
\sigma (\bar{q} q \rightarrow \Phi \Phi) = \frac{2\pi \alpha^{2}_{s}}{9s}
k^3 \,,
\end{equation}
\begin{equation}
\sigma (gg \rightarrow \Phi \Phi) = \frac{\pi \alpha^{2}_{s}}{s}
(\frac{15k}{16} + \frac{51kM^2}{8s} +  \frac{9M^2}{2s^2} (s - M^2)
\ln{(\frac{1-k}{1+k})}) \,.
\end{equation}
where $k = (1 - \frac{4M^2}{s})^{\frac{1}{2}}$. At LHC the main
contribution ($\geq 95$ percent) for the production of scalar octets
comes from the gluon annihilation into two scalar octets $gg
\rightarrow \Phi \Phi$. The scalar octets decay into two gluons that
leads to the four-jet events at LHC. Therefore the single signature
of the scalar octets at LHC are the four-jet events. The main background
comes from QCD four-jet events.
The cross section for the scalar octet production is typically
$O(10^{-4})$ of the standard
QCD two-jet cross section and it is $O(10^{-2})$ of the four-jet QCD
background. The preliminary conclusion is that at LHC for $L_t =
10^{5}$ $pb^{-1}$ it would be possible to discover the scalar octets
with the mass $M \leq 900 $ Gev.

\paragraph{Search for Kaluza-Klein states.}

As it is well known new degrees of freedom are required in any attempt
of unification of the electroweak and strong interactions with gravity.
Among these attempts only superstring theory is known to provide
a consistent description of quantum gravity.
Strings predict two kinds of new degrees of freedom:

(i) Superheavy oscillation modes whose characteristic scale is given
by the inverse
string tension $(\alpha^{'})^{-\frac{1}{2}} \sim 10^{18}$ Gev. These
states are important at very short distances of the order of Planck
length and modify the ultraviolet behavior of gravitational
interactions.

(ii). States associated to the internal compactified
space whose presence is required from the fact that superstring
theory in flat space is anomaly free only in ten dimensions.
Usually, the size of the internal space is also made too small
to give any observable effect in particle accelerators.

However it is possible to construct superstring models having
one or two large internal dimensions at a scale accessible
to LHC \cite{60}. The presence of such large dimensions is motivated by
superstring theory with perturbative breaking of supersymmetry \cite{61}
and their size is inversely proportional to the scale of supersymmetry
breaking which must be  of order of the electroweak scale in order to
protect the gauge hierarchy. In contrast to field theoretical
expectations, string theory allows the existence of such large
dimension(s) consistently with perturbative unification of low-energy
couplings in a class of models based on orbifold compactifications
\cite{62}. Properties of string models with perturbative breaking of
supersymmetry, were studied in the case of minimal embedding of the
standard  model \cite{63}. The main signature of the large extra
dimension(s) in these constructions is the appearance of a tower of
excitations for the gauge bosons and Higgs bosons with the same
gauge quantum numbers. The Kaluza-Klein(KK) states are the
straightforward consequence of all models with compactified dimensions
and they have masses:
\begin{equation}
m^2_n = m^2_0 +\frac{\vec{n}^2}{R^2} \,,
\end{equation}
where R denotes the common radius of the  D large internal
dimensions, $\vec{n}$ is a D dimensional vector and $m_0$ denotes
for R-independent contributions coming from the electroweak
symmetry breaking. All massive KK-states are unstable and they decay
into quarks and leptons with a lifetime $O(10^{-26})$ sec when the size
of the compact dimension(s) is $O(1)Tev^{-1}$. Present
experimental limits have been obtained from an analysis of the effective
four-fermion operators which arise from the exchange of the massive
KK modes \cite{65}. In orbifold models the current limits are
$R^{-1} \geq 185$ Gev for one large extra dimension, while $R^{-1} \geq
1.4$ Tev, 1.1 Tev, 1 Tev for two large dimensions in the case of
$Z_{3}$, $Z_{4}$ and $Z_{6}$ orbifolds respectively \cite{65}.

Among the KK-excitations of different spins, the easiest way to detect
at LHC are the vectors with the quantum numbers of the electroweak
$SU_L(2) \otimes U(1)$ gauge bosons. The most efficient way of
observing KK states in proton-proton collisions is to identify charged
leptons $l^{\pm}$ in the final state. The main background comes from the
Drell-Yan process $pp \rightarrow l^+l^- + X$ with $l = e, \mu$.
In many models with large compactified dimension(s) due to accidental
suppression of the effective coupling constant of the massive
$SU(2)$ vector excitatins, only the $U(1)$ vector excitations
couple to leptons. These states have masses given by equation (243) and
they couple with fermions through the effective interaction:
\begin{equation}
g^{'*}(p)\bar{\psi}^k \gamma_{\mu}(v_k + a_k
\gamma_{5})\psi^{k}B^{*\mu}_{n} \,,
\end{equation}
where k labels the different species of fermions, and $g^{'*}(p)$ is
an effective coupling constant at scale $p$. The interaction (244) leads
to rates of $N^{*}_n$ decays into fermions:
\begin{equation}
\Gamma(B^{*}_n \rightarrow f\bar{f}) = (g^{'*}(m_n)^2
\frac{m_n}{12\pi}C_f(v^2_f + a^2_f) \,,
\end{equation}
while the corresponding interaction with their scalar superpartners
$\tilde{f}_{R,L}$ lead to the decay rates
\begin{equation}
\Gamma(B^{*}_n \rightarrow \bar{\tilde{f}}_{R(L)} \tilde{f}_{R(L)})
= (g^{'*}(m_n))^2 \frac{m_n}{48\pi}C_f(v_f \pm a_f)^2 \,,
\end{equation}
where $C_f = 1$ or 3 for colour singlets ot triplets, respectively.
The total width is:
\begin{equation}
\Gamma_{n} = \frac{5}{8\pi}(g^{'*})^2m_n \,.
\end{equation}
The total cross section for the production of the KK-excitations
$B^{*}_n$ is given by \cite{60}
\begin{equation}
\sigma_t = \sum _{quarks}\int_{0}^{\sqrt{s}}\,dM\, \int_{\ln{M/\sqrt{s}}}
^{\ln{\sqrt{s}/M}} \,dy\, q_q(y,M)S_q(y,M) \,,
\end{equation}
where
\begin{equation}
    g_q(y,M)  = \frac{M}{18\pi}  x_ax_b[f^{(p)}_q(x_a,M)f^{(p)}_
{\bar{q}}(x_b,M) +f^{(p)}_{\bar{q}}(x_a,M)f^{(b)}(x_b,M)] \,,
\end{equation}
and
\begin{equation}
S_q(y, M) = (g^{'*})^4 \frac{1}{N} \sum _{|\vec n | < R\sqrt{s}}
\frac{(v^2_q + a^2_q)(v^2_l + a^2_l)}{(M^2 - m^2_n)^2 +
\Gamma^{2}_{n}m^2_m} \,,
\end{equation}
where $x_a = \frac{M}{\sqrt{s}} e^y$, $x_b = \frac{M}{\sqrt{s}} e^{-y}$,
factor N comes from $Z_N$ orbifold projection. The main background comes
from the Drell-Yan production mechanism. The main conclusion of ref.
\cite{60} is that at LHC for $L_t = 10^{5}$ $pb^{-1}$ it would be possible
to obtain a bound $R^{-1} \geq 4.5$Tev on the compactification radius.

\paragraph{Compositeness}

A composite structure for quarks would appear in the form deviations
from the standard QCD expectations at high transverse momenta, where
valence quark scattering dominates. It is expected  that it would
be possible to discover the quark compositeness scale with
$\Lambda_{c} \le 10$ Tev. The best way to look for the lepton
compositeness effects is the study of lepton pair production
at large dilepton invariant masses. It would be possible to discover
the lepton compositeness for $\Lambda_{lept} \le 20$ Tev \cite{2}.

\paragraph{Nonstandard Higgs bosons}

Many Higgs doublet model where each Higgs doublet couples with its
own quark with relatively big Yukawa coupling constant has been
considered in ref. \cite{65}. For nonsmall Yukawa coupling constants
the main reaction for the production of the Higgs doublets
corresponding to the first and the second generations is quark-antiquark
fusion. The phenomenology of the Higgs doublets corresponding to
the third generation is very similar to the phenomenology of the
model with two Higgs doublets. The cross section for the quark-antiquark
fusion in quark-parton model in the approximation of the infinitely
narrow resonances is given by the standard formula
\begin{eqnarray}
&&\sigma(AB \rightarrow H_{q_iq_j} + X) =
\frac{4\pi^{2}\Gamma(H_{q_iq_j} \rightarrow \bar{q}_iq_j)}{9sM_H}
\int^1_{\frac{M^2_H}{s}}\frac{dx}{x} \\ \nonumber
&&[\bar{q}_{Ai}(x,\mu)q_{Bj}(x^{-1}M^2_Hs^{-1}, \mu) +
q_{Aj}(x,\mu)\bar{q}_{Bj}(x^{-1}M^2_Hs^{-1}, \mu)] \,.
\end{eqnarray}
Here $\bar{q}_{Ai}(x,\mu)$ and $q_{Aj}(x,\mu)$ are parton distributions
of the antiquark $\bar{q}_i$ and quark $q_j$ in hadron A at the
normalization point $\mu \sim M_H$ and the  $\Gamma(H_{q_iq_j}
\rightarrow \bar{q}_iq_j)$ is the hadronic decay width of the Higgs
boson into quark-antiquark pair. For the Yukawa Lagrangian
\begin{equation}
L_Y = h_{q_iq_j}\bar{q}_{Li}q_{Rj}H_{q_iq_j} + h.c.\,,
\end{equation}
the hadronic decay width for massless quarks is
\begin{equation}
\Gamma(H_{q_iq_j} \rightarrow \bar{q}_iq_j) =
\frac{3M_H h^2_{q_iq_j}}{16\pi}\,.
\end{equation}

The value of the renormalization point $\mu$ have been choosen equal to
the mass $M_H$ of the corresponding Higgs boson. The variation of the
renormalization point $\mu$ in the interval $0.5M_H - 2M_H$ leads to
the variation of cross section less than 50 percent. In considered
models there are Higgs bosons which couple both with
down quarks and leptons so the best signature
is the search for the electrically neutral Higgs boson decays
into $e^+e^-$ or $\mu^{+} \mu^{-}$ pairs. For the charged Higgs bosons
the best way to detect them is to look for their decays into charged
leptons and neutrino. The Higgs doublets which couple with up quarks
in model with massless neutrino do not couple with leptons so the only
way to detect them is the search for the resonance type structure
in the distribution of the  dijet cross section on the dijet invariant
mass as in the case of all Higgs bosons, since in the considered models all
Higgs bosons decay mainly into quark-antiquark pairs that leads at the
hadron level to additional dijet events. However the accuracy of the
determination of the dijet cross section is $O(1))$ percent so it would
be not so easy to find stringent bound on the Higgs boson mass by the
measurement of dijet differential cross section at LHC.  In considered
many Higgs doublet model, due to the smallness of the vacuum expectation
values of the Higgs doublets corresponding to the u, d, s and c quarks,
after electroweak symmetry breaking the mass splitting inside the Higgs
doublets is small, so in such models the search for neutral Higgs boson
decaying into lepton pair is in fact the search for the corresponding
Higgs isodoublet. The main background in the search for neutral
Higgs bosons through their decays into lepton pair is the
Drell-Yan process which is under control. The main conclusion of
the ref.\cite{65} is that at LHC for $L_t = 10^{5}$ $pb^{-1}$ and
for the corresponding Yukawa coupling constant $h_{Y} = 1$
it would be possible to detect
such Higgs bosons with the masses up to $4.5 - 5$ Tev.

\paragraph{Search for vector-like fermions.}

Vector-like fermions are characterized by having their left- and
right-handed components transforming in the same way under the
gauge symmetry group. Therefore their mass terms
$\bar{\psi}_L \psi_R$ are not forbidden by any symmetry. As a
consequence their masses are unbounded and they decouple when
they are taken to infinity. The standard model does not need vector-like
fermions and the models with additional vector-like fermions are
not very popular at present. Vector-like fermions decay by the exchange
of standard electroweak gauge and Higgs bosons , $W^{\pm}$, $Z$, $h$,
with all three being comparable in size \cite{66}. It is possible
to add to standard model extra vector-like fermions with the
following quantum numbers:

a. Down singlet D quark.

b. Up singlet U quark.

c. Up down quark doublet $(U,D)$

d. Singlet charged lepton E.

e. Singlet neutral lepton N.

f. Neutral charged lepton doublet $(N, E)$.

The production cross section are similar to those of standard fermions.

Vector-like quarks decay mainly into $Q \rightarrow W q_i,\, Z q_i,\,
h q_i$ modes, while vector-like leptons decay mainly into
$L \rightarrow Wl,\,Zl,\,hl$ modes. For vector-like quarks
the branchings obey the approximate rule: $W/Z/h \approx 2/1/1$.
The signatures from vector-like quark pair production at LHC
with their subsequent decays are: 6 jet events, 2 leptons + 4 jets
events and 4 leptons plus 2 jets events. The signatures from
the vector-like lepton pair production with
their subsequent decays are: 6 lepton events, 4 lepton plus two jets
events, 2 leptons plus 4 jets events.

\newpage

\section{Conclusion}

LHC can test the structure of many theories at Tev
scale. LHC will be able to discover Higgs boson and low
energy broken supersymmetry with squark and gluino masses up to
(2 - 2.5) Tev. Also there is nonzero chance to find something
new ($Z^{'}$-bosons, $W^{'}$-bosons, leptoquarks...) at LHC.
At any rate after LHC we will know the basic elements of the
matter structure at Tev region.

We are indebted to our colleagues from INR theoretical department
for useful discussions. The research described in this publication
was made possible in part by Award No RP1-187 of the U.S. Civilian
Research and Development Foundation for the Independent States of the
Former Soviet Union(CRDF).

\end{document}